\documentclass[lettersize, onecolumn, journal,noadjust]{IEEEtran}
\usepackage{amsmath,amsfonts}
\usepackage{algorithmic}
\usepackage{algorithm}
\usepackage{array}
\usepackage[caption=false,font=normalsize,labelfont=sf,textfont=sf]{subfig}
\usepackage{textcomp}
\usepackage{stfloats}
\usepackage{url}
\usepackage{verbatim}
\usepackage{graphicx}
\usepackage{cite}

\usepackage{amssymb}
\usepackage{dsfont}
\usepackage{float}
\usepackage{graphicx}
\newtheorem{lemma}{Lemma}

\newtheorem{definition}{Definition}
\newtheorem{proposition}{Proposition}
\newtheorem{theorem}{Theorem}
\newtheorem{remark}{Remark}
\newtheorem{corollary}{Corollary}

\newtheorem{condition}{Condition}
\usepackage{xcolor}
\usepackage{mathrsfs}
\usepackage{caption}
\usepackage{multirow}
\usepackage{makecell}
\usepackage{mathtools} 
\usepackage[ colorlinks = true,              linkcolor = blue, urlcolor  =
blue,              citecolor = red,              anchorcolor = green,
]{hyperref}

\usepackage [autostyle, english = american]{csquotes}
\MakeOuterQuote{"}

\usepackage{amsmath}

\begin{document}

\title{Channel Coding for Gaussian Channels with Multifaceted Power Constraints}

\author{%
  \begin{tabular}[t]{@{}c@{\hspace{6em}}c@{}}
    Adeel Mahmood & Aaron B.~Wagner \\
    Radio Systems Research & School of Electrical and Computer Engineering\\
    Nokia Bell Labs & Cornell University
  \end{tabular}
}



\maketitle

\begin{abstract}
Through refined asymptotic analysis based on the normal approximation, we study how higher-order coding performance depends on the mean power as well as on finer statistics of the input power. We introduce a multifaceted power model in which the expectation of an arbitrary (but finite) number of arbitrary 
functions of the normalized average power is constrained. The framework generalizes existing models, recovering the standard maximal and expected power constraints and the recent mean and variance constraint as special cases. Under
certain growth and continuity assumptions on the functions, our main theorem gives an exact characterization of the minimum average error probability for Gaussian channels as a function of the first- and second-order coding rates. The converse proof reduces the code design problem to minimization over a compact (under the Prokhorov metric) set of probability distributions, characterizes the extreme points of this set and invokes the Bauer's maximization principle. Our results for the multifaceted power model serve as more precise benchmarks for practical modulation schemes with multiple amplitude levels, probabilistic shaping and nonuniform constellation geometries.
\end{abstract}

\begin{IEEEkeywords}
Channel coding, Gaussian channels, second-order coding rate, power
constraint.
\end{IEEEkeywords}

\section{Introduction}
For Additive White Gaussian Noise (AWGN) channels with a maximal power constraint,  
\begin{align}
    \lim_{n \to\infty} P_e(n, R, \Gamma) = \begin{cases}
        0 & \text{if $R < C(\Gamma)$} \\
        1 & \text{if $R > C(\Gamma)$},
    \end{cases} \label{first-order}
\end{align}
where $P_e(n, R, \Gamma)$ is the minimum average error probability
over all block codes with rate $R$, blocklength $n$ and power
$\Gamma$, and $C(\Gamma)$ is the well-known capacity of the channel:
\begin{equation}
C(\Gamma) = \frac{1}{2} \log \left(1 + \frac{\Gamma}{N}\right).
\label{eq:capacity}
\end{equation}
First-order asymptotic results in channel coding such as $(\ref{first-order})$ provide relatively little insight into the design of practical communication systems which operate at finite blocklengths. Shannon~\cite{review11} derived nonasymptotic upper and lower bounds on the minimum average error probability with
application to the error-exponent regime, in which the rate $R$ is a constant independent of $n$. For rates approaching capacity, refined second- and third-order asymptotic studies based on the normal approximation are often used instead to shed additional light into the effects of finite/short blocklengths on the achievable coding performance \cite{strassen, 5290292, Polyanskiy2010,review13}. For example, a second-order refinement of $(\ref{first-order})$ is \cite[Theorem 5]{5290292}
\begin{align}
    \lim_{n \to\infty} P_e\left(n, C(\Gamma) + \frac{r}{\sqrt{n}}, \Gamma\right) = \Phi\left(\frac{r}{\sqrt{V(\Gamma)}} \right), \label{second-order}
\end{align}
where $V(\Gamma)$ is the channel dispersion, $\Phi(\cdot)$ is
the standard Normal CDF, and $r$ is called the \emph{second-order coding rate (SOCR)}. We also have a refined asymptotic expansion of the maximum rate $R^*(n,\epsilon, \Gamma)$ as a function of a fixed error probability $\epsilon \in (0, 1)$ as   
\begin{align}
    R^*(n,\epsilon, \Gamma) = C(\Gamma) + \frac{\sqrt{V(\Gamma)} \Phi^{-1}(\epsilon)}{\sqrt{n}}  + \frac{1}{2} \frac{\log n}{n} + O\left(\frac{1}{n}\right), \label{third-orderc}
\end{align}
which includes a third-order $\log n / n$ term as well \cite{7056434}, \cite[Theorem 54]{5452208}. Upper and lower bounds on the $O(1/n)$ term in $(\ref{third-orderc})$ can be found in \cite[Theorem 5]{4thorderkostina}. Such refined asymptotic results can lead to conclusions which differ from those based on first-order asymptotic results only. For example, feedback does not increase the capacity (first-order) but can increase the second-order coding rate \cite{9099482}; separate source-channel coding is first-order but not second-order
optimal \cite{kostina_JSCC}. For additional such examples, see~\cite[p. 4598]{7156144}.

While significant attention has been given to obtaining refined results with respect to the blocklength $n$ as in $(\ref{second-order})$ and $(\ref{third-orderc})$, the dependence of these results on the signaling power (or channel input cost) characteristics is rather coarse, i.e., the dependence is only captured by a single parameter $\Gamma$ in $(\ref{first-order})$-$(\ref{third-orderc})$. One might say that such results are first-order only with respect to the channel input power statistics. Common forms of cost constraints in channel coding \cite{review12} have been the maximal cost constraint specified by $c(\mathbf{X}) \leq \Gamma$ almost surely, or the expected cost constraint specified by $\mathbb{E}\left [ c(\mathbf{X}) \right] \leq \Gamma$, where $c(\cdot)$ is a cost function expressed as 
\begin{align}
    c(\mathbf{X}) \coloneqq \frac{1}{n} \sum_{i=1}^n c(X_i). \label{nletterextension}
\end{align}
With only a single parameter $\Gamma$ included in the cost model, previous results in channel coding have characterized the optimal coding performance whose dependence on the signaling power statistics is limited. For example, while the single-parameter maximal cost constraint model is a reasonable approximation for constant-envelope modulations such as PSK, more sophisticated schemes like QAM are nonconstant-envelope and use multiple amplitude levels. In shaped QAM \cite{shaped_QAM}, the symbol (constellation point) probabilities are made nonuniform, often with ring-dependent PMFs tied to symbol energy, so two transmitters with the same nominal power budget $\Gamma$ 
can have different codeword-power fluctuations. The classical AWGN finite-blocklength benchmark $(\ref{third-orderc})$ does not capture those finer statistics: Polyanskiy \cite{Polyanskiy2010} showed that equal-power and maximal-power constraints have the same first- and second-order asymptotics, so a benchmark indexed only by $\Gamma$ cannot distinguish a shell-like input from a multiring, shaped one. Hence, more refined cost models are needed to better benchmark sophisticated signaling schemes whose finite-blocklength performance depends not just on the mean power $\Gamma$.

Recent works \cite{mahmood2024channelcodingmeanvariance}, \cite{mahmood2024improvedchannelcodingperformance}, \cite{mahmood2025channelcodinggaussianchannels} 
give one such refinement by extending the cost model to include both the mean and variance parameters. Specifically, subject to 
\begin{align}
\begin{split}
    \mathbb{E}\left [c(\mathbf{X}) \right] &\leq  \Gamma\\
    \text{Var}\left ( c(\mathbf{X}) \right) &\leq \frac{V}{n},
\end{split}
     \label{43v} 
\end{align}
the works characterize the optimal coding performance as a function of both $\Gamma$ and $V$ as 
\begin{align}
    R^*(n, \epsilon, \Gamma, V) = C(\Gamma) + \frac{r^*(\epsilon, \Gamma, V)}{\sqrt{n}} + o\left(\frac{1}{\sqrt{n}}\right), \label{mvsecond-order}
\end{align}
where an expression for $r^*(\epsilon, \Gamma, V)$ can be found in \cite[(25)]{mahmood2024channelcodingmeanvariance}\footnote{A slight refinement of \cite[(25)]{mahmood2024channelcodingmeanvariance} can be obtained by applying \cite[Lemma 3]{mahmood2025channelcodinggaussianchannels} to it.}. By additionally accounting for the variance of the channel input cost, the second-order term 
\begin{align*}
    \frac{r^*(\epsilon, \Gamma, V)}{\sqrt{n}}
\end{align*}
in the asymptotic expansion $(\ref{mvsecond-order})$ is a refinement of the second-order term 
\begin{align*}
    \frac{\sqrt{V(\Gamma)} \Phi^{-1}(\epsilon)}{\sqrt{n}}
\end{align*}
in the asymptotic expansion $(\ref{third-orderc})$, which was allowed to depend on $\Gamma$ only.

Notably, the first-order terms in both $(\ref{third-orderc})$ and $(\ref{mvsecond-order})$ are the same (thus the strong converse holds) because the variance constraint enforces concentration around the mean $\Gamma$. In the absence of the variance constraint in $(\ref{43v})$, the strong converse does not hold \cite{7055296}. Hence, higher-order terms have delicate dependence on the statistics of the channel input power, even if the channel input power concentrates around the mean $\Gamma$ leaving the first-order term $C(\Gamma)$ unchanged. In this paper, we aim to characterize this dependence further by including an arbitrary (but finite) number of cost parameters $\Gamma, \Gamma_1, \ldots, \Gamma_k$ in the cost model. We focus on AWGN channels only and take the cost function to be $c(x) = x^2$, with its $n$-letter extension given as in $(\ref{nletterextension})$. The parameter $\Gamma$ corresponds to the mean constraint, as in $(\ref{43v})$, but each of the subsequent parameters corresponds to a moment-type constraint of the form 
\begin{align}
    \mathbb{E}\left [ f_i \left( \frac{1}{\sqrt{n}} \sum_{i=1}^n \left [ c(X_i) - \Gamma \right ]\right) \right ] \leq \Gamma_i,
\end{align}
specified by a function $f_i : \mathbb{R} \to [0, \infty)$. We will show that when the functions $(f_1, \ldots, f_k)$ are chosen to enforce \emph{uniform upper-tail cost concentration}\footnote{defined in $(\ref{g13})$ and $(\ref{f1})$ later}, then the strong converse holds and only the second-order term depends on $(\Gamma, \Gamma_1, \ldots, \Gamma_k)$, similar to the result in $(\ref{mvsecond-order})$ for the mean and variance cost constraint. A simple sufficient condition on the functions $f_1, \ldots, f_k$ to enforce uniform upper-tail cost concentration is also given (see Condition \ref{reg_cond2}). Our main result expresses the minimum average error probability as a function of the coding rate, similar to the form given in $(\ref{second-order})$.      

To encapsulate, we consider a multifaceted cost framework given by  
\begin{align}
    \mathbb{E}\left [  \widetilde{c}(\mathbf{X}) \right ] &\leq 0 \label{cw1}\\
    \mathbb{E}\left [ f_i\left(\widetilde{c}(\mathbf{X}) \right) \right] &\leq \Gamma_i \quad \text{ for } i = 1, \ldots, k,  \label{cw2}
\end{align}
where we call
\begin{align}
    \widetilde{c}(\mathbf{X}) := \frac{1}{\sqrt{n}} \sum_{i=1}^n \left [ c(X_i) - \Gamma \right] = \sqrt{n} (c(\mathbf{X}) - \Gamma) \label{ncd}
\end{align}
the \emph{normalized cost deviation}.
The cost framework of $(\ref{cw1})$ and $(\ref{cw2})$ generalizes several cost formulations from prior works:
\begin{itemize}
\item To recover the maximal cost constraint, let $k = 1, \Gamma_1 = 0$ and 
\begin{align*}
    f_1(u) = \begin{cases}
        0 & \text{if $u \leq 0$} \\
        u & \text{if $u > 0$},
    \end{cases}
\end{align*}
in $(\ref{cw1})$ and $(\ref{cw2})$ to obtain 
\begin{align}
    \mathbb{P}\left( c(\mathbf{X}) \leq \Gamma \right) &= 1. \label{c2as}
\end{align}
    \item To approximate the mean and variance cost constraint given in $(\ref{43v})$, let $k=1$, $\Gamma_1 = V$ and $f_1(u) = u^2$ in $(\ref{cw1})$ and $(\ref{cw2})$ to obtain 
\begin{align}
\begin{split}
    \mathbb{E}\left [c(\mathbf{X}) \right] &\leq \Gamma \\
    \mathbb{E}\left [ \left( c(\mathbf{X}) - \Gamma \right)^2 \right] &\leq \frac{V}{n}.
\end{split}
\label{mb}
\end{align}
The difference is that (\ref{mb}) penalizes quadratic deviations
from $\Gamma$ instead of from the mean of $c(\mathbf{X})$. 
But this difference is immaterial because optimal codes achieve
equality in the first constraint; see the discussion in Section~\ref{sec:discussion}.
\item To recover an expectation-only constraint, choose $k = 0$.
\item To recover an excess cost probability constraint, let $k = 1, \Gamma_1 = \delta > 0$ and 
\begin{align*}
    f_1(u) = \begin{cases}
        0 & u \leq 0\\
        1 & u > 0
    \end{cases}
\end{align*}
in $(\ref{cw1})$ and $(\ref{cw2})$ to obtain 
\begin{align}
    \mathbb{E}\left [c(\mathbf{X}) \right] &\leq \Gamma \label{c1e}\\
    \mathbb{P}\left( c(\mathbf{X}) > \Gamma \right) &\leq \delta. \label{c2e}
\end{align}
\end{itemize}
The analog of $(\ref{c2e})$ is called the \emph{excess distortion probability} in source coding, which is a common performance metric in lossy compression \cite{Marton:Exponent,6145679,Zhou:Two,Truong:Joint:Feedback,Venkataramanan:SPARC:Excess,Zhou:SR,Zhou:GrayWyner,Kelly:WZ,Zhong:Joint,Zhong:Continuous,Weissman:Exponent:Universal,Iriyama:General,Hen:Trellis,Venkataramanan:Feedforward,Shi:Semantic}. However, excess cost probability is a less common performance metric in channel coding (but see \cite{Hori:Overrun,Hori:Overrun:ISITA,Nishiara:Delivery}).

The remainder of the paper is organized as follows. The next section contains the problem formation, statement of the main result, and
further discussion. Preliminaries to the proof are given
in Sec.~\ref{sec:preliminaries}. Sec. ~\ref{sec:converse} contains the converse proof and
Sec.~\ref{sec:achievability} contains the achievability proof.
Concluding remarks are given in Sec.~\ref{sec:conclusion}.
Technical proofs are relegated to the appendices.

\section{Formulation and Result Statement}
\label{sec:main}

Let $\mathcal{P}(\mathbb{R}^n)$ denote the set of all Borel probability measures on $\mathbb{R}^n$. 
We write $\mathbf{x} = (x_1, \ldots, x_n)$ to denote a vector and $\mathbf{X} = (X_1, \ldots, X_n)$ to denote a random vector in $\mathbb{R}^n$.  For any $\mu \in \mathbb{R}$ and $\sigma^2 > 0$, let $\mathcal{N}(\mu, \sigma^2)$ denote the Gaussian distribution with mean $\mu$ and variance $\sigma^2$. 
The AWGN channel $W(\cdot|x) = \mathcal{N}(x, N)$ models the relationship between the channel input $\mathbf{X}$ and output $\mathbf{Y}$ over $n$ channel uses as $\mathbf{Y} = \mathbf{X} + \mathbf{Z}$, where $\mathbf{Z} \sim \mathcal{N}(\mathbf{0}, N \cdot \mathbf{I}_n)$ represents independent and identically distributed (i.i.d.) Gaussian noise with variance $N > 0$. The noise vector $\mathbf{Z}$ is independent of the input $\mathbf{X}$.  

Let the cost function $c : \mathbb{R} \to [0, \infty )$ be given by $c(x) = x^2$. For a channel input sequence $\mathbf{x} \in \mathbb{R}^n$, 
\begin{align*}
    c(\mathbf{x}) =  \frac{1}{n} \sum_{i=1}^n c(x_i) = \frac{||\mathbf{x}||^2}{n}.
\end{align*}

For $\Gamma > 0$, the capacity-cost function of the channel $W$ with cost threshold $\Gamma$ is defined as  
\begin{align}
    C(\Gamma) := \max_{P: \mathbb{E}_P[c(X)] \leq \Gamma} I(P, W), \label{defcc}
\end{align}
where 
\begin{align*}
    \mathbb{E}_P\left [c(X) \right ] = \int_{\mathbb{R}} x^2 dP(x).  
\end{align*}
In the Gaussian case, this is well-known to be as given in~(\ref{eq:capacity})~\cite[(9.17)]{Cover2006}.

With a blocklength $n$ and a fixed rate $R > 0$, let $\mathcal{M}_R = \{1, \ldots, \lceil \exp(nR) \rceil \}$ denote the message set. Let $M \in \mathcal{M}_R$ denote the random message drawn uniformly from the message set. An $(n, R)$ channel code consists of an encoder $\operatorname{enc} : \mathcal{M}_R \to \mathbb{R}^n$ and a decoder $\operatorname{dec}: \mathbb{R}^n \to \mathcal{M}_R$.    

Recall the definition of $\widetilde{c}(\mathbf{X})$ in $(\ref{ncd})$. We consider random channel codes for which the admissible distributions of the channel input satisfy an average-deviation constraint and $k$ additional moment-type constraints (for any $k \geq 1$) specified by $f \coloneqq (f_1, \ldots, f_k)$ and $\mathbf{\Gamma} \coloneqq (\Gamma, \Gamma_1, \ldots, \Gamma_k)$ as follows: 
\begin{align}
    \mathbb{E}\left [  \widetilde{c}(\mathbf{X}) \right ] &\leq 0 \label{c1}\\
    \mathbb{E}\left [ f_i\left(\widetilde{c}(\mathbf{X}) \right) \right] &\leq \Gamma_i \quad \text{ for } i = 1, \ldots, k,  \label{c2}
\end{align}
where each $f_i : \mathbb{R} \to [0, \infty)$ and $(\Gamma, \Gamma_1, \ldots, \Gamma_k) \in (0, \infty) \times [0, \infty)^k$. In $(\ref{c1})$ and $(\ref{c2})$, the expectation is with respect to both the random message $M$ and the codebook randomness. In other words, $\mathbf{X} = \operatorname{enc}(M)$ is a function of both $M$ and any private randomness used by the channel encoder. Throughout,
we allow for randomized codes.

\begin{definition}
    Given an $(n, R)$ channel code $(\operatorname{enc}, \operatorname{dec})$, the distribution $\overline{P}$ of the channel input induced by the code is given by 
    \begin{align}
    \overline{P}(A) := \frac{1}{\lceil \exp(nR) \rceil} \sum_{m=1}^{\lceil \exp(nR) \rceil} \mathbb{P}\left(\text{enc}(m) \in A \right) \label{ohwelli}
\end{align}
    for any Borel subset $A \subset \mathbb{R}^n$, where the probability $\mathbb{P}(\cdot)$ in the RHS of $(\ref{ohwelli})$ is with respect to the codebook randomness.   
    \label{induced_def}
\end{definition}

\begin{definition}[Admissible distributions]\label{def:PnfG}
For each $n \geq 1$, let $\mathcal{P}_{n, f, \mathbf{\Gamma}} := \left \{ \overline{P} \in P(\mathbb R^n):
        \mathbf{X} \sim \overline{P} \text{ satisfies } (\ref{c1}) \text{ and } (\ref{c2}) \right\}.$
\label{defcostcons}
\end{definition}
Throughout the paper, we will assume that $f$ and $\mathbf{\Gamma}$ are such that $\mathcal{P}_{n, f, \mathbf{\Gamma}}$ is nonempty for every integer $n \geq 1$. Some results in the paper will additionally require one or both of the following two conditions:  
\begin{condition}
$f = (f_1, \ldots, f_k)$ and $\mathbf{\Gamma} = (\Gamma, \Gamma_1, \ldots, \Gamma_k)$ are such that
each $f_i : \mathbb{R} \to [0, \infty)$ is a Borel measurable, lower–semicontinuous function and
$$\mathbf{\Gamma} \in (0, \infty) \times [0,\infty)^k.$$
\label{reg_cond}
\end{condition}
\begin{condition}
    At least one of the functions in 
$f = (f_1, \ldots, f_k)$ is eventually nondecreasing and diverges to infinity, i.e., there exists $i$ and $x_0$  such that $f_i$ is nondecreasing on $[x_0,\infty)$ and $\lim_{x \to \infty} f_i(x) = \infty$.  
\label{reg_cond2}
\end{condition}
Condition \ref{reg_cond2} ensures uniform upper-tail concentration in the following sense: 
\begin{align}
    \sup_{\overline{P} \in \mathcal{P}_{n, f, \mathbf{\Gamma}} } \mathbb{P}_{\overline{P}}\left( \widetilde{c}(\mathbf{X}) >  \sqrt{n} a_n \right) \to 0 \text{ as } n \to \infty  \label{g13}
\end{align}
or equivalently, 
\begin{align}
    \sup_{\overline{P} \in \mathcal{P}_{n, f, \mathbf{\Gamma}} } \mathbb{P}_{\overline{P}} \left( c(\mathbf{X}) > \Gamma + a_n \right) \to 0 \text{ as } n \to \infty \label{f1}
\end{align}  
for all $o(1)$ sequences $a_n$ such that $a_n > 0$ and $\sqrt{n} a_n \to \infty$. 

\begin{definition}
    We define $\mathcal{C}_{n, R, f, \mathbf{\Gamma}}$ as the class of $(n, R)$ channel codes such that $\overline{P} \in \mathcal{P}_{n, f, \mathbf{\Gamma}}$, where $\overline{P}$ is the channel
    input distribution induced\footnote{in the sense of Definition \ref{induced_def}} by the code and $\mathcal{P}_{n, f, \mathbf{\Gamma}}$ is defined in Definition \ref{defcostcons}. 
    \label{defcodeset}
\end{definition}

The following theorem is the main result of the paper.

\begin{theorem}
 Let $W(\cdot|\mathbf{x}) = \mathcal{N}(\mathbf{x}, N \mathbf{I}_n)$. Given any
 $f = (f_1, \ldots, f_k)$ and $\mathbf{\Gamma} = (\Gamma, \Gamma_1, \ldots, \Gamma_k)$, let $\mathcal{C}_{n, R, f, \mathbf{\Gamma}}$ be the set of $(n, R)$ channel codes as defined in Definition \ref{defcodeset}. Let $R = C(\Gamma) + r/\sqrt{n}$ for any fixed real number $r$. Assume that $f$ and $\mathbf{\Gamma}$ satisfy Condition \ref{reg_cond} and Condition \ref{reg_cond2}. For a channel code $\mathscr{C} \in \mathcal{C}_{n, R, f, \mathbf{\Gamma}}$, let $\epsilon(\mathscr{C})$ denote its
average error probability. Then 
\begin{align}
    \lim_{n \to \infty} \inf_{\mathscr{C} \in
         \mathcal{C}_{n,
     R, f, \mathbf{\Gamma}}} \epsilon(\mathscr{C}) = \inf_{ \substack{P_{U} \in \mathcal{U}_{f, \mathbf{\Gamma}}\\|\text{supp}(P_U)| \leq k+2} } \mathbb{E}_{P_U} \left [  \Phi\left( \frac{r}{\sqrt{V(\Gamma)}} - \frac{C'(\Gamma)U}{\sqrt{V(\Gamma)}}  \right) \right ], \label{m1-b}
\end{align}
where
\begin{align}
    \mathcal{U}_{f, \mathbf{\Gamma}} := \left \{ P \in \mathcal{P}(\mathbb{R}) : \mathbb{E}_{P}[U] \leq 0, \mathbb{E}_{P}[f_i(U)] \leq \Gamma_i \text{ for } i = 1, \ldots, k \right \}.
\end{align}
\label{main_thm}
\end{theorem}

The converse half of the proof is provided in Section~\ref{sec:converse} and the
achievability half is provided in Section~\ref{sec:achievability}.

\subsection{Discussion}
\label{sec:discussion}

The result $(\ref{m1-b})$ is a generalization of 
\begin{align}
    \lim_{n \to\infty} P_e\left(n, C(\Gamma) + \frac{r}{\sqrt{n}}, \Gamma\right) = \Phi\left(\frac{r}{\sqrt{V(\Gamma)}} \right), \label{n2...4}
\end{align}
where the generalization holds for an achievability scheme that uses random codewords drawn from a mixture of $k + 2$ uniform distributions on $(n-1)$-spheres of radii $R_1, \ldots, R_{k + 2}$, where $R_i = O(\sqrt{n})$ and $|R_i - R_j| = O(1)$. For a maximal power constraint, drawing codewords from a single $(n-1)$-sphere of radius $\sqrt{n\Gamma}$ achieves $(\ref{n2...4})$ and is optimal. For the mean and variance power constraint, at least two spheres of different radii are necessary while three are sufficient to achieve the optimal error probability \cite[Section IV]{mahmood2024improvedchannelcodingperformance}; see also \cite[Fig. 1 \& Fig. 2]{mahmood2025channelcodinggaussianchannels}. For the multifaceted power model with $k + 1$ constraints as in $(\ref{c1})$-$(\ref{c2})$, we show in subsection \ref{main_thm_achiev_proof} that $k + 2$ spheres are  sufficient for achieving the optimal error probability $(\ref{m1-b})$.

We use a mixture distribution over $k + 2$ spheres in the random codebook construction for attaining the optimal error probability $(\ref{m1-b})$. It is worth mentioning that this is structurally similar to the fact that the capacity-achieving distribution for an $N$-dimensional additive
Gaussian noise channel with an amplitude-constrained input is supported on finitely many concentric shells \cite{dytso2019}, \cite{dytso2020}. But despite the structural similarity of the distributions, the two results address different optimization problems. The problem in this paper is an operational problem over a scalar AWGN channel: among all $(n, R)$ channel codes such that the code-induced codeword distribution satisfies the constraints $(\ref{c1})$-$(\ref{c2})$, we characterize the minimum average error probability at rates $R = C(\Gamma) + r/\sqrt{n}$, in the limit as $n \to \infty$. In our achievability construction, each mass point $u_j$, where $j = 1, \ldots, k + 2$, of an arbitrary distribution $P_U$ in $(\ref{m1-b})$ is mapped to a power shell with radius $\sqrt{n\Gamma_j}$, where $\Gamma_j = \Gamma + u_j/\sqrt{n}$. In contrast, the problem in \cite{dytso2019}, \cite{dytso2020} is a mutual information maximization problem for an $N$-dimensional vector Gaussian noise channel with an amplitude constraint 
$||X^N||_2 \leq \operatorname{A}$. Here, $N$ is the dimension of one vector channel use, whereas $n$ in our paper is the number of channel uses of a scalar AWGN channel. The aforementioned maximum mutual information is the "information capacity" \cite[(7.1)]{Cover2006} of the amplitude-constrained vector channel, and it describes the maximum asymptotically achievable rate when the $N$-dimensional vector Gaussian channel is used many times with each channel use super-symbol $X^N$ constrained as $||X^N|| \leq \operatorname{A}$. This is different from the block-level constraints $(\ref{c1})$-$(\ref{c2})$ considered in our paper.

The uniform upper-tail concentration property $(\ref{f1})$ can be seen as a counterpart of the variance constraint 
\begin{align}
    \text{Var}\left( c(\mathbf{X}) \right) \leq \frac{V}{n} \label{f2}
\end{align}
used in \cite{mahmood2025channelcodinggaussianchannels,mahmood2024channelcodingmeanvariance,mahmood2024improvedchannelcodingperformance}. Unlike $(\ref{f2})$, however, $(\ref{f1})$ only constrains deviations above the threshold $\Gamma$. Furthermore, while $(\ref{f2})$ implies uniform upper-tail concentration in $(\ref{f1})$, the property $(\ref{f1})$ or even a two-sided concentration property
\begin{align}
    \sup_{\overline{P} \in \mathcal{P}_{n, f, \mathbf{\Gamma}} } \mathbb{P}_{\overline{P}} \left( \left |c(\mathbf{X}) - \Gamma \right | >  a_n \right) \to 0 \text{ as } n \to \infty, \label{fbn1}
\end{align}  
for all $o(1)$ sequences $a_n$ such that $\sqrt{n} a_n \to \infty$, does not enforce $(\ref{f2})$. Therefore, the present paper allows a larger set of allowed cost behaviors while still retaining the structural consequences such as the strong converse and finite second-order coding rate proved under the mean and variance cost formulation.

We now examine Theorem~\ref{main_thm} in several special cases. Recall that $R = C(\Gamma) + r/\sqrt{n}$.
Under the maximal cost formulation in $(\ref{c2as})$, it follows from Theorem~\ref{main_thm} that the average error probability satisfies
\begin{align}
    \lim_{n \to \infty} \inf_{\mathscr{C} \in \mathcal{C}_{n,
     R, f, \mathbf{\Gamma}}} \epsilon(\mathscr{C}) 
    &= \inf_{\substack{U:\\
    \mathbb{P}\left(U > 0 \right) =0 \\|\text{supp}(U)| \leq 3} } \mathbb{E} \left [  \Phi\left( \frac{r}{\sqrt{V(\Gamma)}} - \frac{C'(\Gamma)U}{\sqrt{V(\Gamma)}}  \right) \right ]\\
    &= \Phi\left(\frac{r}{\sqrt{V(\Gamma)}} \right),
\end{align}
which is exactly the second-order coding performance characterized under the a.s. cost constraint~\cite[Theorem~5]{5290292} (cf.~(\ref{second-order})).

Under the mean-and-variance cost formulation in $(\ref{mb})$, it follows from Theorem~\ref{main_thm} that the average error probability satisfies \begin{align}
    \lim_{n \to \infty} \inf_{\mathscr{C} \in \mathcal{C}_{n,
     R, f, \mathbf{\Gamma}}} \epsilon(\mathscr{C}) 
    &= \inf_{\substack{U:\\ \mathbb{E}[U] \leq 0\\
    \mathbb{E}[U^2] \leq V \\|\text{supp}(U)| \leq 3} } \mathbb{E} \left [  \Phi\left( \frac{r}{\sqrt{V(\Gamma)}} - \frac{C'(\Gamma)U}{\sqrt{V(\Gamma)}}  \right) \right ].\label{mvspecial}
\end{align}
It is straightforward to see that the inequality constraint $\mathbb{E}[U] \leq 0$ in $(\ref{mvspecial})$ can be replaced with the equality constraint $\mathbb{E}[U] = 0$. Hence, the RHS of $(\ref{mvspecial})$ can be rewritten as 
\begin{align}
    \inf_{\substack{U:\\ \mathbb{E}[U] = 0\\
    \operatorname{Var}(U) \leq V \\|\text{supp}(U)| \leq 3} } \mathbb{E} \left [  \Phi\left( \frac{r}{\sqrt{V(\Gamma)}} - \frac{C'(\Gamma)U}{\sqrt{V(\Gamma)}}  \right) \right ]
    &= \inf_{\substack{\Pi:\\ \mathbb{E}[\Pi] = \frac{r}{\sqrt{V(\Gamma)}}\\
    \operatorname{Var}\left(\Pi\right) \leq \frac{C'(\Gamma)^2 V}{V(\Gamma)} \\|\text{supp}(\Pi)| \leq 3} } \mathbb{E} \left [  \Phi\left(\Pi  \right) \right ], 
\end{align}
which matches the result for the mean and variance cost constraint \cite{mahmood2025channelcodinggaussianchannels}.   

In practice, one might be more concerned with deviations of
$c(\mathbf{X})$ above $\Gamma$ than below. Consider the case of
$k = 1$ and
\begin{equation}
    f_1(u) = \begin{cases}
        0 & \text{if $u < 0$} \\
        u^2 & \text{if $u \ge 0$}.
    \end{cases}
\end{equation}
Theorem~\ref{main_thm} characterizes the limiting error probability for this case:
\begin{align}
    \lim_{n \to \infty} \inf_{\mathscr{C} \in \mathcal{C}_{n,
     R, f, \mathbf{\Gamma}}} \epsilon(\mathscr{C}) 
    &= \inf_{\substack{U:\\ \mathbb{E}[U] \leq 0\\
    \mathbb{E}[f_1(U)] \leq \Gamma_1 \\|\text{supp}(U)| \leq 3} } \mathbb{E} \left [  \Phi\left( \frac{r}{\sqrt{V(\Gamma)}} - \frac{C'(\Gamma)U}{\sqrt{V(\Gamma)}}  \right) \right ]. \label{prev_res_one_sided}
\end{align}

For an expectation-only constraint ($k = 0$), the strong
converse does not hold and~\cite[Thm.~77]{Polyanskiy2010}
\begin{align}
    \lim_{n \to \infty} \inf_{\mathscr{C} \in \mathcal{C}_{n,
     R, f, \mathbf{\Gamma}}} \epsilon(\mathscr{C}) 
    &= 0.
\end{align}
Theorem~\ref{main_thm} does not directly apply when $k = 0$, but an upper bound on the error probability in this case is given by the RHS of $(\ref{prev_res_one_sided})$ 
for any $\Gamma_1$. Then letting $\Gamma_1 \rightarrow \infty$ gives
\begin{align}
    \lim_{n \to \infty} \inf_{\mathscr{C} \in \mathcal{C}_{n,
     R, f, \mathbf{\Gamma}}} \epsilon(\mathscr{C}) 
    & \le \inf_{\substack{U:\\ \mathbb{E}[U] \leq 0\\
    |\text{supp}(U)| \leq 3} } \mathbb{E} \left [  \Phi\left( \frac{r}{\sqrt{V(\Gamma)}} - \frac{C'(\Gamma)U}{\sqrt{V(\Gamma)}}  \right) \right ],
\end{align}
which is known to be zero for all $r$~\cite[p.~1509]{mahmood2024channelcodingmeanvariance}.

Finally, consider the problem with an \emph{excess cost probability} constraint:
\begin{equation}
P(\widetilde{c}(\mathbf{X}) > \gamma) \le \delta.
\label{eq:excesscost}
\end{equation}
which corresponds to $k = 1$, $\Gamma_1 = \delta$, and 
\begin{equation}
f_1(u) = \begin{cases}
    1 & \text{if $u > \gamma$} \\
    0 & \text{if $u \le \gamma$}.
\end{cases}
\end{equation}
The choice of $f_1$ above does not satisfy Condition \ref{reg_cond2}, but $f_1$ can be approximated by
\begin{equation}
    f^{(\alpha)}_1(u) = \begin{cases}
            1 + \alpha(u - \gamma) & \text{if $u > \gamma$} \\
            0 & \text{if $u \le \gamma$}
        \end{cases}
\end{equation}
for small $\alpha$, and for this choice, 
Condition \ref{reg_cond2} is satisfied for any $\alpha > 0$.
In Appendix~\ref{app:excess}, it is shown that 
\begin{align}
    \lim_{\alpha \rightarrow 0} \inf_{\substack{U:\\ \mathbb{E}[U] \leq 0\\ 
            \mathbb{E}[f^{(\alpha)}_1(U)] \le \delta \\
    |\text{supp}(U)| \leq 3} } \mathbb{E} \left [  \Phi\left( \frac{r}{\sqrt{V(\Gamma)}} - \frac{C'(\Gamma)U}{\sqrt{V(\Gamma)}}  \right) \right ] =
      (1-\delta) 
    \Phi\left( \frac{r}{\sqrt{V(\Gamma)}} - \frac{C'(\Gamma)\gamma}{\sqrt{V(\Gamma)}}  \right).
\end{align}

\section{Preliminaries to the Proof}
\label{sec:preliminaries}

For any Borel-measurable function $g : \mathbb{R}^n \to \mathbb{R}$ and any $P \in \mathcal{P}(\mathbb{R}^n)$, we write $g_{\#} P$ for the pushforward measure, i.e., $(g_{\#} P)(B) = (P \circ g^{-1})(B)$ for any Borel set $B \subset \mathbb{R}$. For any subset $\mathcal{P} \subset \mathcal{P}(\mathbb{R}^n)$, we define $g_{\#} \mathcal{P} \subset \mathcal{P}(\mathbb{R})$ as the pushforward set: 
\begin{align*}
    g_{\#} \mathcal{P} := \left \{ g_{\#} P : P \in \mathcal{P}  \right\}.  
\end{align*} 

Let $\chi^2_n(\lambda)$ denote the noncentral chi-squared distribution with $n$ degrees of freedom and noncentrality parameter $\lambda$. If two random variables $X$ and $Y$ have the same distribution, we write $X \stackrel{d}{=} Y$. We will write $\log$ to denote logarithm to the base $e$ and $\exp(x)$ to denote $e^x$. If $P \in \mathcal{P}(\mathbb{R}^n)$ is an $n$-fold product distribution induced by some $P' \in \mathcal{P}(\mathbb{R})$, then we write 
\begin{align}
    P(\mathbf{x}) = \prod_{i=1}^n P'(x_i) = P'(\mathbf{x}) \label{abuse}
\end{align}
where the second equality above involves some abuse of notation. Given any $P \in \mathcal{P}(\mathbb{R}^n)$ and a conditional probability distribution or a channel $W(\cdot|\cdot)$, we use $P \circ W$ to denote the joint probability distribution and $PW$ to denote the induced output distribution, i.e., if $(\mathbf{X}, \mathbf{Y}) \sim P \circ W$, then $\mathbf{X} \sim P$ and $\mathbf{Y} \sim PW$. If $W$ is a scalar channel, we write $W(\mathbf{y} | \mathbf{x}) = \prod_{i=1}^n W(y_i | x_i)$ similar to the notation in $(\ref{abuse})$.

\begin{definition}
    We use $P^*$ to denote the capacity-cost-achieving distribution in $(\ref{defcc})$ and $Q^* = P^*W$ to denote the induced output distribution. We define 
    \begin{align*}
    \nu_{x} &:= \text{Var}\left( \log \frac{W(Y|x)}{Q^*(Y)} \right),\quad  \text{ where } Y \sim W(\cdot|x),\\
    V(\Gamma) &:= \int_{\mathbb{R}} \nu_x P^*(x) dx.
\end{align*}
\end{definition}

\begin{lemma} For an AWGN channel $W(\cdot|x) = \mathcal{N}(x, N)$, we have $P^* = \mathcal{N}(0, \Gamma)$ and $Q^* = \mathcal{N}(0, \Gamma + N)$. 
For any $x \in \mathbb{R}$ and $Y \sim W(\cdot|x)$, 
    \begin{align}
   \mathbb{E}\left [ \log \frac{W(Y|x)}{Q^*(Y)} \right] &= C(\Gamma) - C'(\Gamma) \left(\Gamma - c(x) \right) \label{equsentimes} \\
    C'(\Gamma) &= \frac{1}{2(\Gamma + N)} \notag \\
    \nu_x &= \frac{\Gamma^2 + 2x^2N}{2\left(N + \Gamma \right)^2} \notag \\
    V(\Gamma) &= \frac{\Gamma^2 + 2\Gamma N }{2\left(N + \Gamma \right)^2}. \notag
\end{align}
\label{oftusedlemma}
\end{lemma}
\begin{IEEEproof} The fact that $P^* = \mathcal{N}(0, \Gamma)$ can be found in \cite[(9.17)]{Cover2006}. The other assertions follow from elementary calculus.
\end{IEEEproof}

In deriving optimal first- and second-order coding rates subject to $(\ref{c1})$ and $(\ref{c2})$, we will show that the block-level constraints in $(\ref{c1})$ and $(\ref{c2})$ can be mapped to the set of scalar distributions given in $(\ref{scalardist})$. Then using Gaussian approximation, the optimal second-order performance can be expressed as a function similar in form to the one given in $(\ref{be})$.      

\begin{lemma}
    Given any constants $\kappa_1, \kappa_2$,
$f = (f_1, \ldots, f_k)$ and $\mathbf{\Gamma} = (\Gamma, \Gamma_1, \ldots, \Gamma_k)$ such that
\begin{align}
    \mathcal{U}_{f, \mathbf{\Gamma}} := \left \{ P \in \mathcal{P}(\mathbb{R}) : \mathbb{E}_{P}[U] \leq 0,  \mathbb{E}_{P}[f_i(U)] \leq \Gamma_i \text{ for } i = 1, \ldots, k \right \} \label{scalardist}
\end{align}
is nonempty, define the function 
 \begin{align}
    g(r) = \inf_{\substack{P_{U} \in \mathcal{U}_{f, \mathbf{\Gamma}}\\ |\text{supp}(P_U)| \leq k + 2 }} \mathbb{E}_{P_U} \left [  \Phi\left( \kappa_1 \,r - \kappa_2 \,U  \right) \right ]. \label{be} 
\end{align}
Thus the function $g$ is Lipschitz continuous with Lipschitz constant $\kappa_1 / \sqrt{2\pi}.$ 
\label{contrlemma}
\end{lemma}
\begin{IEEEproof} 
    The proof is given in Appendix \ref{contrlemma_proof}.
\end{IEEEproof}

\section{Converse}
\label{sec:converse}

The starting point of the converse proof is the following
result relating the error probability of a code over two channels.
Many similar results have appeared in the literature 
(e.g.,~\cite[Thm.~26]{5452208}, ~\cite[Thm.~27]{5452208}, \cite[(42)]{8012458}, \cite[Lemma~15]{9099482}). The version given below uses \cite[Thm. 26]{5452208} and Neyman-Pearson threshold inequality (see, e.g., \cite[(102)]{5452208}), while restricting the set of channel input distributions $\overline{P}$ to be in $\mathcal{P}_{n, f, \mathbf{\Gamma}}$.  

\begin{lemma}
Fix any two channels $W$ and $\widetilde{W}$. 
If a code $\mathscr{C} \in \mathcal{C}_{n, R, f, \mathbf{\Gamma}}$ has average error probability $\epsilon' \in (0,1)$ when used over the channel $\widetilde{W}$ and average error probability at most $\epsilon \in (0, 1)$ when used over the channel $W$, then for any real number $\gamma > 0$ and integer $n > 0$, 
\begin{align*}
    -\log (1-\epsilon') &\leq n \gamma - \log \left [ \left( \inf_{\overline{P} \in \mathcal{P}_{n, f, \mathbf{\Gamma}} } (\overline{P} \circ W) \left( \log  \frac{W(\mathbf{Y}|\mathbf{X})}{\widetilde{W}(\mathbf{Y}|\mathbf{X})} \leq n \gamma \right) - \epsilon  \right)^+\right]. 
\end{align*}
\label{gen_convo_lemmi}
\end{lemma}

\begin{remark}
Lemma \ref{gen_convo_lemmi} is closely related to the the meta-converse (specifically \cite[Thm. 27]{5452208}), but it differs in three respects. First, the infimum over input distributions is over a restricted set $\mathcal{P}_{n, f, \mathbf{\Gamma}}$ as a result of the multifaceted power constraint. Second, the lemma is derived from
the more general comparison result \cite[Thm. 26]{5452208}, which
allows for an input-dependent auxiliary channel $\widetilde{W}$ rather than
only an input-independent auxiliary output distribution. Third, Lemma \ref{gen_convo_lemmi} substitutes $\beta_{1 - \epsilon}(P, Q)$ with the lower bound $\beta_{1 - \epsilon}(P, Q) \geq \frac{1}{\eta}\left(\mathbb{P}_P\left( \frac{dP}{dQ} \leq \eta \right) - \epsilon \right)$ that holds for all $\eta > 0$. 
\end{remark}

For an example where an auxiliary channel instead of an output distribution is used to prove a converse result, see \cite[p. 4601]{7156144}. For our purposes, we will apply Lemma \ref{gen_convo_lemmi} to the case where $\widetilde{W}(\mathbf{Y}|\mathbf{X})$ is a fixed Gaussian distribution
$q \in \mathcal{P}(\mathbb{R}^n)$ that does not depend on $\mathbf{X}$.
In this case, the following two results characterizing the
distribution of $\log W(\mathbf{Y}|\mathbf{X})/q(\mathbf{Y})$ will
prove useful.

\begin{lemma}
    Let $W(\cdot|\mathbf{x}) = \mathcal{N}(\mathbf{x}, N \mathbf{I}_n )$ and $q = \mathcal{N}(\mathbf{0}, (\Gamma + N)\mathbf{I}_n)$. Let $(\mathbf{X}, \mathbf{Y}) \sim \overline{P} \circ W$ for any $\overline{P} \in \mathcal{P}(\mathbb{R}^n)$. Then 
    \begin{align}
        \log \frac{W(\mathbf{Y}|\mathbf{X})}{q(\mathbf{Y})} \stackrel{d}{=} n C(\Gamma)  + \frac{nS}{2\Gamma} - \sum_{i=1}^n \frac{\Gamma}{2(N + \Gamma)} \left( \tilde{Z}_i - \frac{\sqrt{N S}}{\Gamma}  \right)^2, \label{condionS=s}
    \end{align}
    where $S = \frac{||\mathbf{X}||^2}{n}$ and $\tilde{Z}_i$'s are i.i.d. $\mathcal{N}(0, 1)$ and independent of $S$.
    \label{log_dist_eq}
\end{lemma}
\begin{IEEEproof}
The proof is given in Appendix~\ref{log_dist_eq_proof}.
\end{IEEEproof}

Conditioning $(\ref{condionS=s})$ on any fixed realization $S = s$, the RHS of $(\ref{condionS=s})$ can be expressed as a sum of i.i.d. random variables, which allows the central limit theorem to be applied. Lemma \ref{lemma_gen_lower_bnd_wo_opt} below gives the result obtained from such an argument.

\begin{lemma}
      Let $W(\cdot|\mathbf{x}) = \mathcal{N}(\mathbf{x}, N \mathbf{I}_n )$ and $q = \mathcal{N}(\mathbf{0}, (\Gamma + N)\mathbf{I}_n)$. Let $(\mathbf{X}, \mathbf{Y}) \sim \overline{P} \circ W$ for any $\overline{P} \in \mathcal{P}(\mathbb{R}^n)$. Then for every $\gamma > 0$ and integer $n > 0$, 
      \begin{align}
        (\overline{P} \circ W) \left( \log  \frac{W(\mathbf{Y}|\mathbf{X})}{q(\mathbf{Y})} \leq n \gamma \right) \geq \mathbb{E}_{P_S} \left [ \Phi\left( \frac{\sqrt{2n}(N + \Gamma)( \gamma - C(\Gamma))}{\sqrt{\Gamma^2 + 2 N S}} + \frac{\sqrt{n}(\Gamma - S)}{\sqrt{2}\sqrt{\Gamma^2 + 2 N S}}  \right)  \right] - \frac{15^{\frac{3}{4}}}{\sqrt{n}},  
        \label{eq:lemma_gen_lower_bnd_wo_opt}
      \end{align}

      where $S = \frac{||\mathbf{X}||^2}{n}$. \label{lemma_gen_lower_bnd_wo_opt}
\end{lemma}
\begin{IEEEproof}
The proof is given in Appendix~\ref{lemma_gen_lower_bnd_wo_opt_proof}.
\end{IEEEproof}

The lower bound $(\ref{eq:lemma_gen_lower_bnd_wo_opt})$ simplifies the dependence on $\overline{P}$ only through the induced law $P_S$ of $S = \frac{||\mathbf{X}||^2}{n} $. Hence, an optimization over any feasible set of $\overline{P} \in \mathcal{P}(\mathbb{R}^n)$ can be replaced by an optimization over a corresponding feasible set of $P_S \in \mathcal{P}([0, \infty))$, simplifying the  subsequent analysis. Indeed, if $\overline{P}$ is contained in 
$\mathcal{P}_{n, f, \mathbf{\Gamma}}$, then the induced $P_S$
is contained in
      \begin{align*}
      \mathcal{S}_{n,f,\mathbf{\Gamma}} &\coloneqq \left \{ c_{\# }\overline{P} : \overline{P} \in \mathcal{P}_{n, f, \mathbf{\Gamma}}  \right\}\\    &\,= \left \{ P \in \mathcal{P}([0, \infty)) : \mathbb{E}_{P}[S] \leq \Gamma, \mathbb{E}_{P}[f_i(\sqrt{n}(S  - \Gamma))] \leq \Gamma_i \text{ for } i = 1, \ldots, k \right \}
      \end{align*}
      where $c : \mathbb{R}^n \to [0, \infty)$
is defined as $c(\mathbf{X}) = \frac{||\mathbf{X}||^2}{n}$. 
The set $\mathcal{S}_{n,f,\mathbf{\Gamma}}$ is evidently convex.
It is also compact, a fact that will be useful in the sequel.

\begin{lemma}
Given any
	$f = (f_1, \ldots, f_k)$ and $\mathbf{\Gamma} = (\Gamma, \Gamma_1, \ldots, \Gamma_k)$ satisfying Condition \ref{reg_cond}, the set $\mathcal{S}_{n, f, \mathbf{\Gamma}}$ is compact under the Prokhorov metric~\cite[p.~72]{Billingsley:Convergence2}.
    \label{sncompact}
\end{lemma}
\begin{IEEEproof}
The proof is given in Appendix~\ref{sncompactproof}.
\end{IEEEproof}

The infimum over the left-hand side of (\ref{eq:lemma_gen_lower_bnd_wo_opt}) over 
$\overline{P} \in \mathcal{P}_{n, f, \mathbf{\Gamma}}$ is 
evidently lower bounded by the infimum of the right-hand side 
over $P_S \in \mathcal{S}_{n,f,\mathbf{\Gamma}}$. Using the
compactness of $\mathcal{S}_{n,f,\mathbf{\Gamma}}$ one
can show that the latter is achieved.

\begin{lemma}
Given any
$f = (f_1, \ldots, f_k)$ and $\mathbf{\Gamma} = (\Gamma, \Gamma_1, \ldots, \Gamma_k)$ satisfying Condition \ref{reg_cond}, for every $\gamma > 0$ and integer $n > 0$, 
    \begin{align*}
        &\inf_{P_{S} \in \mathcal{S}_{n, f, \mathbf{\Gamma}}} \mathbb{E}_{P_S} \left [ \Phi\left( \frac{\sqrt{2n}(N + \Gamma)( \gamma - C(\Gamma))}{\sqrt{\Gamma^2 + 2 N S}} + \frac{\sqrt{n}(\Gamma - S)}{\sqrt{2}\sqrt{\Gamma^2 + 2 N S}}  \right)  \right] \\
        &\quad \quad \quad \quad \quad \quad \quad \quad = \min_{P_{S} \in \mathcal{S}_{n, f, \mathbf{\Gamma}}} \mathbb{E}_{P_S} \left [ \Phi\left( \frac{\sqrt{2n}(N + \Gamma)( \gamma - C(\Gamma))}{\sqrt{\Gamma^2 + 2 N S}} + \frac{\sqrt{n}(\Gamma - S)}{\sqrt{2}\sqrt{\Gamma^2 + 2 N S}}  \right)  \right].
    \end{align*}
\label{lemminac}
\end{lemma}
\begin{IEEEproof}
The proof is given in Appendix~\ref{lemminacproof}. 
\end{IEEEproof}

Define a function $\phi_{n, \gamma} : [0, \infty) \to [0, 1]$ as 
\begin{align*}
    \phi_{n, \gamma}(s) &=   \Phi\left( \frac{\sqrt{2n}(N + \Gamma)( \gamma - C(\Gamma))}{\sqrt{\Gamma^2 + 2 N s}} + \frac{\sqrt{n}(\Gamma - s)}{\sqrt{2}\sqrt{\Gamma^2 + 2 N s}}  \right).
\end{align*}

Define the continuous, linear functional $\mathcal{L}(P) = \mathbb{E}_P\left [ \phi_{n, \gamma}(S) \right]$. Consider  
\begin{align}
    \min_{P_S \in \mathcal{S}_{n, f, \mathbf{\Gamma}}} \mathcal{L}(P_S). \label{4cvb}
\end{align}
In $(\ref{4cvb})$, we have a minimization of a continuous, linear functional over a compact and convex set. From Bauer's maximization principle \cite[7.69]{AliprantisBorder2006}, the minimum is attained at one of the extreme points of the set $\mathcal{S}_{n, f, \mathbf{\Gamma}}$. From \cite[Theorem~2.1]{winkler1988extreme}, the extreme points of $\mathcal{S}_{n, f, \mathbf{\Gamma}}$ are probability distributions with at most $k+2$ point masses. Hence, we have the following lemma:

\begin{lemma}
    Given any
$f = (f_1, \ldots, f_k)$ and $\mathbf{\Gamma} = (\Gamma, \Gamma_1, \ldots, \Gamma_k)$ satisfying Condition \ref{reg_cond}, for every $\gamma > 0$ and integer $n > 0$, 
    \begin{align*}
        &\inf_{P_{S} \in \mathcal{S}_{n, f, \mathbf{\Gamma}}} \mathbb{E}_{P_S} \left [ \Phi\left( \frac{\sqrt{2n}(N + \Gamma)( \gamma - C(\Gamma))}{\sqrt{\Gamma^2 + 2 N S}} + \frac{\sqrt{n}(\Gamma - S)}{\sqrt{2}\sqrt{\Gamma^2 + 2 N S}}  \right)  \right] \\
        &\quad \quad \quad \quad \quad \quad \quad = \min_{\substack{P_{S} \in \mathcal{S}_{n, f, \mathbf{\Gamma}}\\|\text{supp}(P_S)| \leq k+2 }} \mathbb{E}_{P_S} \left [ \Phi\left( \frac{\sqrt{2n}(N + \Gamma)( \gamma - C(\Gamma))}{\sqrt{\Gamma^2 + 2 N S}} + \frac{\sqrt{n}(\Gamma - S)}{\sqrt{2}\sqrt{\Gamma^2 + 2 N S}}  \right)  \right].
    \end{align*}
\label{gvb}
\end{lemma}

Combining Lemma~\ref{lemma_gen_lower_bnd_wo_opt} and Lemma~\ref{gvb} gives the following corollary.

\begin{corollary}
 Let $W(\cdot|\mathbf{x}) = \mathcal{N}(\mathbf{x}, N \mathbf{I}_n )$ and $q = \mathcal{N}(\mathbf{0}, (\Gamma + N)\mathbf{I}_n)$. Let $(\mathbf{X}, \mathbf{Y}) \sim \overline{P} \circ W$. Given any
$f = (f_1, \ldots, f_k)$ and $\mathbf{\Gamma} = (\Gamma, \Gamma_1, \ldots, \Gamma_k)$ satisfying Condition \ref{reg_cond}, for every $\gamma > 0$ and integer $n > 0$,
\begin{align*}
        \inf_{\overline{P} \in \mathcal{P}_{n, f, \mathbf{\Gamma}}}(\overline{P} \circ W) \left( \log  \frac{W(\mathbf{Y}|\mathbf{X})}{q(\mathbf{Y})} \leq n \gamma \right) \geq \min_{\substack{P_{S} \in \mathcal{S}_{n, f, \mathbf{\Gamma}}\\|\text{supp}(P_S)| \leq k+2 }} \mathbb{E}_{P_S} \left [ \Phi\left( \frac{\sqrt{2n}(N + \Gamma)( \gamma - C(\Gamma))}{\sqrt{\Gamma^2 + 2 N S}} + \frac{\sqrt{n}(\Gamma - S)}{\sqrt{2}\sqrt{\Gamma^2 + 2 N S}}  \right)  \right] - \frac{15^{\frac{3}{4}}}{\sqrt{n}}.
      \end{align*}
\label{genlem3pt}
\end{corollary}

Our focus is on large $n$, and in this asymptotic regime the
bound in Corollary~\ref{genlem3pt} can be reduced to the following.

\begin{lemma}
    Let $W(\cdot|\mathbf{x}) = \mathcal{N}(\mathbf{x}, N \mathbf{I}_n )$ and $q = \mathcal{N}(\mathbf{0}, (\Gamma + N)\mathbf{I}_n)$. Let $(\mathbf{X}, \mathbf{Y}) \sim \overline{P} \circ W$. Then for every $r \in \mathbb{R}$ and for every $f$ and $\mathbf{\Gamma}$ satisfying Condition \ref{reg_cond} and Condition \ref{reg_cond2},
    \begin{align*}
        \lim_{n \to \infty} \inf_{\overline{P} \in \mathcal{P}_{n, f, \mathbf{\Gamma}}}(\overline{P} \circ W) \left( \log  \frac{W(\mathbf{Y}|\mathbf{X})}{q(\mathbf{Y})} \leq n C(\Gamma) + r \sqrt{n} \right) &\geq \inf_{ \substack{P_{U} \in \mathcal{U}_{f, \mathbf{\Gamma}}\\|\text{supp}(P_U)| \leq k+2} } \mathbb{E}_{P_U} \left [  \Phi\left( \frac{\sqrt{2}(N + \Gamma)r}{\sqrt{\Gamma^2 + 2 N \Gamma}} - \frac{U}{\sqrt{2}\sqrt{\Gamma^2 + 2 N \Gamma}}  \right) \right ]\\
    &= \inf_{ \substack{P_{U} \in \mathcal{U}_{f, \mathbf{\Gamma}}\\|\text{supp}(P_U)| \leq k+2} } \mathbb{E}_{P_U} \left [  \Phi\left( \frac{r}{\sqrt{V(\Gamma)}} - \frac{C'(\Gamma)U}{\sqrt{V(\Gamma)}}  \right) \right ],
      \end{align*}
      where
      \begin{align}
    \mathcal{U}_{f, \mathbf{\Gamma}} := \left \{ P \in \mathcal{P}(\mathbb{R}) : \mathbb{E}_{P}[U] \leq 0, \mathbb{E}_{P}[f_i(U)] \leq \Gamma_i \text{ for } i = 1, \ldots, k \right \}.
\end{align}
      \label{asymplemma}
\end{lemma}
\begin{IEEEproof}
The proof is given in Appendix~\ref{asymplemmaproof}.
\end{IEEEproof}

We are now in a position to prove the impossibility half of Theorem~\ref{main_thm}.  

\begin{proposition}
    In the setup of Theorem~\ref{main_thm}, we have
\begin{align}
    \liminf_{n \to \infty} \inf_{\mathscr{C} \in
         \mathcal{C}_{n,
     R, f, \mathbf{\Gamma}}} \epsilon(\mathscr{C}) \ge \inf_{ \substack{P_{U} \in \mathcal{U}_{f, \mathbf{\Gamma}}\\|\text{supp}(P_U)| \leq k+2} } \mathbb{E}_{P_U} \left [  \Phi\left( \frac{r}{\sqrt{V(\Gamma)}} - \frac{C'(\Gamma)U}{\sqrt{V(\Gamma)}}  \right) \right ].
\end{align}
\label{main_prop_converse}
\end{proposition}

\subsection{Proof of Proposition~\ref{main_prop_converse} \label{main_prop_converse_proof}}

The proof is essentially substituting the result of Lemma \ref{asymplemma} into Lemma \ref{gen_convo_lemmi} followed by using the continuity property from Lemma \ref{contrlemma}. For any channel code $\mathscr{C} \in \mathcal{C}_{n, R, f, \mathbf{\Gamma}}$ with $R = C(\Gamma) + r/\sqrt{n}$, we have $|\mathcal{M}_R| = \exp\left(n C(\Gamma) + r \sqrt{n} \right)$. 
We apply Lemma~\ref{gen_convo_lemmi} with $\gamma = C(\Gamma) + r'/\sqrt{n}$, for any $r' < r$ and $W(\mathbf{Y}|\mathbf{X}) =
q(\mathbf{Y})$, where $q = \mathcal{N}(0,(\Gamma + N) \mathbf{I}_n)$,
to obtain 
    \begin{align*}
        \log |\mathcal{M}_R| &\leq n C(\Gamma) + \sqrt{n} r'  - \log \left [ \left( \inf_{\overline{P} \in \mathcal{P}_{n, f, \mathbf{\Gamma}} } (\overline{P} \circ W) \left( \log  \frac{W(\mathbf{Y}|\mathbf{X})}{q(\mathbf{Y})} \leq n \gamma \right) - \epsilon  \right)^+\right]\\
         r \sqrt{n} &\leq  \sqrt{n} r'  - \log \left [ \left( \inf_{\overline{P} \in \mathcal{P}_{n, f, \mathbf{\Gamma}} } (\overline{P} \circ W) \left( \log  \frac{W(\mathbf{Y}|\mathbf{X})}{q(\mathbf{Y})} \leq n \gamma \right) - \epsilon  \right)^+\right]. 
    \end{align*}
    We then have 
    \begin{align}
        \left( \inf_{\overline{P} \in \mathcal{P}_{n, f, \mathbf{\Gamma}} } (\overline{P} \circ W) \left( \log  \frac{W(\mathbf{Y}|\mathbf{X})}{q(\mathbf{Y})} \leq n \gamma \right) - \epsilon  \right)^+ &\leq \exp\left((r' - r) \sqrt{n}\right)
    \end{align}
    which implies that the average probability of error $\epsilon$ is lower bounded as 
    \begin{align}
        \epsilon &\geq \inf_{\overline{P} \in \mathcal{P}_{n, f, \mathbf{\Gamma}} } (\overline{P} \circ W) \left( \log  \frac{W(\mathbf{Y}|\mathbf{X})}{q(\mathbf{Y})} \leq n \gamma \right) - \exp\left((r' - r) \sqrt{n}\right)\\
        &= \inf_{\overline{P} \in \mathcal{P}_{n, f, \mathbf{\Gamma}} } (\overline{P} \circ W) \left( \log  \frac{W(\mathbf{Y}|\mathbf{X})}{q(\mathbf{Y})} \leq n C(\Gamma) + \sqrt{n}r' \right) - \exp\left((r' - r) \sqrt{n}\right).
    \end{align}
Taking the limit as $n \to \infty$ and using the result of Lemma \ref{asymplemma}, we obtain 
\begin{align}
    \liminf_{n \to \infty} \epsilon \geq \inf_{ \substack{P_{U} \in \mathcal{U}_{f, \mathbf{\Gamma}}\\|\text{supp}(P_U)| \leq k+2} } \mathbb{E}_{P_U} \left [  \Phi\left( \frac{r'}{\sqrt{V(\Gamma)}} - \frac{C'(\Gamma)U}{\sqrt{V(\Gamma)}}  \right) \right ]. \label{rfv}
\end{align}
From Lemma \ref{contrlemma}, we have that the RHS of $(\ref{rfv})$ is continuous in the variable $r'$. Since $(\ref{rfv})$ holds for an arbitrary $r' < r$, letting $r' \to r$ establishes the result.

\section{Achievability}
\label{sec:achievability}

The starting point of the achievability proof is the following lemma
that mirrors Lemma~\ref{gen_convo_lemmi}.

\begin{lemma}
Consider an AWGN channel $W$ with noise variance $N > 0$. Given any
$f = (f_1, \ldots, f_k)$ and $\mathbf{\Gamma} = (\Gamma, \Gamma_1, \ldots, \Gamma_k)$, let $\mathcal{C}_{n, R, f, \mathbf{\Gamma}}$ be the set of $(n, R)$ channel codes as defined in Definition \ref{defcodeset}. For any $n$, $R$ and $\theta$, the minimum average probability of error of channel codes in $\mathcal{C}_{n, R, f, \mathbf{\Gamma}}$ is upper bounded by   
    \begin{align}
        \inf_{\overline{P} \in \mathcal{P}_{n, f, \mathbf{\Gamma}} }( \overline{P} \circ W) \left(\frac{1}{n} \log \frac{W(\mathbf{Y}|\mathbf{X})}{\overline{P}W(\mathbf{Y})} \leq R + \theta \right) + e^{-n \theta}. \label{patanahimujhe}
    \end{align}
    \label{aaron'slemma}
\end{lemma}

\begin{IEEEproof}
The proof can be adapted from the proof of \cite[Lemma 14]{9099482} by (i) replacing controllers with distributions $\overline{P}$ such that $\overline{P} \in \mathcal{P}_{n, f, \mathbf{\Gamma}}$ and (ii) replacing sums with integrals.
\end{IEEEproof}

We shall choose $\overline{P}$ to be uniform over an $(n-1)$-sphere, or a finite mixture of such distributions. The following result characterizes $\overline{P} W$ in the former
case.

\begin{lemma}
    Consider a random vector $\mathbf{Y} = \mathbf{X} + \mathbf{Z}$, where $\mathbf{X}$ and $\mathbf{Z}$ are independent, $\mathbf{X}$ is uniformly distributed on an $(n-1)$-sphere of radius $R$ and $\mathbf{Z} \sim \mathcal{N}(\mathbf{0}, N \mathbf{I}_n)$. Let $Q^{cc}$ denote the PDF of $\mathbf{Y}$. Then 
    \begin{align*}
        Q^{cc}(\mathbf{y}) = \frac{\Gamma\left( \frac{n}{2} \right)}{2 (\pi N)^{n/2}}  \cdot  \exp\left( -\frac{R^2 + ||\mathbf{y}||^2}{2N} \right)  \left( \frac{N}{R||\mathbf{y}||}\right)^{\frac{n}{2}-1}  I_{\frac{n}{2}-1}\left(\frac{R||\mathbf{y}||}{N}\right),
    \end{align*}
    where $I_{\nu}(x)$ denotes the modified Bessel function of the first kind of order $\nu$.  
    \label{QccN}
\end{lemma}

\begin{IEEEproof}
See \cite[Lemma 5]{mahmood2025channelcodinggaussianchannels}. 
\end{IEEEproof}

The distribution $Q^{cc}$ is nearly a multivariate Gaussian
distribution. The next lemma quantifies the discrepancy between
the two distributions. Variants of this result can be found in  \cite[Lemmas 6 and 7]{mahmood2025channelcodinggaussianchannels},  \cite[(425)]{5452208} and \cite[Proposition 2]{7300429}.  
\begin{lemma}
    Consider a random vector $\mathbf{Y} = \mathbf{X} + \mathbf{Z}$, where $\mathbf{X}$ and $\mathbf{Z}$ are independent, $\mathbf{X}$ is uniformly distributed on an $(n-1)$ sphere of radius $\sqrt{n \Gamma}$ and $\mathbf{Z} \sim \mathcal{N}(\mathbf{0}, N \mathbf{I}_n)$. Let $Q^{cc}$ denote the PDF of $\mathbf{Y}$ and let $Q^* = \mathcal{N}(\mathbf{0}, (\Gamma' + N) \mathbf{I}_n)$. Let $\Gamma' = \Gamma + \epsilon$, where $\epsilon \in \mathbb{R}$ is such that $|\epsilon| < \Gamma + N$. Then for any $0 < \Delta < \Gamma + N - |\epsilon|$, sufficiently small $|\epsilon|$ and sufficiently large $n$,   
    \begin{align}
        \sup_{\mathbf{y} \in \mathcal{P}_n^*} \left( \log \frac{Q^{cc}(\mathbf{y})}{Q^*(\mathbf{y})} \right)  \leq \frac{n\epsilon ^2}{4 \Gamma ^2}-\frac{n \epsilon ^3}{3 \Gamma ^3} + O(n\epsilon^4) + O\left(1 \right),  \label{weirdlemma12bnd} 
    \end{align}
where $\mathcal{P}_n^* = \left \{ \mathbf{y} \in \mathbb{R}^n: \Gamma' + N - \Delta  \leq  \frac{||\mathbf{y}||^2}{n} \leq \Gamma' + N + \Delta  \right \}$ and the $O(n\epsilon^4)$ and $O(1)$ terms can be chosen to be independent of $\Delta$.  
\label{slightlybetterlemma}
\end{lemma}
\begin{remark}
The bound $(\ref{weirdlemma12bnd})$ should be interpreted as follows: There exist positive constants $C_1, C_2, \delta$ and $N$ such that whenever $|\epsilon| \leq \delta$ and $n \geq N$, the LHS of $(\ref{weirdlemma12bnd})$ is upper bounded by 
\begin{align*}
    \frac{n\epsilon ^2}{4 \Gamma ^2}-\frac{n \epsilon ^3}{3 \Gamma ^3} + R_1(n, \epsilon) + R_2(n, \epsilon)
\end{align*}
with $|R_1(n, \epsilon)| \leq C_1 n \epsilon^4$ and $|R_2(n, \epsilon)| \leq C_2$. Furthermore, the parameters $\epsilon$ and $\Delta$ in Lemma \ref{slightlybetterlemma} may depend on $n$, in which case they would be $o(1)$ and $O(1)$ sequences, respectively. For example, $\epsilon = \frac{1}{\sqrt{n}}$, $\Delta = \sqrt{\frac{\log n}{n}}$ and $\Delta =\frac{2}{3}(\Gamma + N - |\epsilon|)$ are valid choices. Finally, note that Lemma \ref{slightlybetterlemma} is different from \cite[Proposition 2]{7300429} in that it allows a small $\epsilon$ power-mismatch   between the $Q^{cc}$-inducing spherical input and the $Q^*$-inducing i.i.d. Gaussian input.   
\end{remark}
\begin{IEEEproof}
The proof of Lemma \ref{slightlybetterlemma} is given in Appendix~\ref{slightlybetterlemmaproof}.
\end{IEEEproof}

We are now in a position to prove the achievability half of Theorem~\ref{main_thm}.

\begin{proposition}
    In the setup of Theorem~\ref{main_thm}, we have
\begin{align}
    \limsup_{n \to \infty} \inf_{\mathscr{C} \in
         \mathcal{C}_{n,
     R, f, \mathbf{\Gamma}}} \epsilon(\mathscr{C}) \le
    \inf_{ \substack{P_{U} \in \mathcal{U}_{f, \mathbf{\Gamma}}\\|\text{supp}(P_U)| \leq k+2} } \mathbb{E}_{P_U} \left [  \Phi\left( \frac{r}{\sqrt{V(\Gamma)}} - \frac{C'(\Gamma)U}{\sqrt{V(\Gamma)}}  \right) \right ].
\end{align}
\label{main_prop_achiev}
\end{proposition}

\subsection{Proof of Proposition~\ref{main_prop_achiev} \label{main_thm_achiev_proof}}

Consider 
\begin{align}
    \inf_{P_{U} \in \mathcal{U}_{f, \mathbf{\Gamma}}^{(k)} } \mathbb{E}_{P_U} \left [  \Phi\left( \frac{r}{\sqrt{V(\Gamma)}} - \frac{C'(\Gamma)U}{\sqrt{V(\Gamma)}}  \right) \right ],
\end{align}
where
\begin{align}
    \mathcal{U}_{f, \mathbf{\Gamma}}^{(k)} := \left \{ P \in \mathcal{P}(\mathbb{R}) : |\text{supp}(P_U)| \leq k+2, \mathbb{E}_{P}[U] \leq 0, \mathbb{E}_{P}[f_i(U)] \leq \Gamma_i \text{ for } i = 1, \ldots, k \right \}.
\end{align}

Let $P_U \in \mathcal{U}_{f, \mathbf{\Gamma}}^{(k)}$ be any arbitrary distribution. Let 
\begin{align*}
    P_{U}(u) = \begin{cases}
    p_1 & u = u_1\\
    p_2 & u = u_2\\
     & \vdots \\
    p_{k + 2} & u = u_{k+2}.
    \end{cases}
\end{align*}
For each $j \in \{1,2, \ldots, k + 2 \}$, let
\begin{align}
\Gamma_j = \Gamma + \frac{u_j}{\sqrt{n}}. \label{38n}   
\end{align}
We assume sufficiently large $n$ so that $\Gamma_j > 0$ for all $j \in \{1,2, \ldots, k + 2 \}$. Let $P^*_j$ be the capacity-cost-achieving input distribution for $C(\Gamma_j)$ and $Q_j^*$ be the corresponding optimal output distribution. Thus, $P_j^* = \mathcal{N}(0, \Gamma_j)$ and $Q_j^* = \mathcal{N}(0, \Gamma_j + N)$. Let $Q^{cc}_j$ be the output distribution induced by the input distribution $\text{Unif}(S^{n-1}_{R_j})$, where $S^{n-1}_{R_j}$ is
the surface of the $n$-ball with radius $R_j = \sqrt{n \Gamma_j}$.

\textbf{Achievability Scheme:} Let the random channel input $\mathbf{X}$ be such that with probability $p_j$, $\mathbf{X} \sim \text{Unif}(S^{n-1}_{R_j})$. Denoting the distribution of $\mathbf{X}$ by $\overline{P}$, we can write 
\begin{align}
    \overline{P} = \sum_{j=1}^{k+2} p_j \cdot \text{Unif}(S^{n-1}_{R_j}). \label{53vn}
\end{align}
The output distribution of $\mathbf{Y}$ induced by $\overline{P} \circ W$ is 
\begin{align*}
    \overline{P}W(\mathbf{y}) &=  \sum_{j=1}^{k+2} p_j Q_j^{cc}(\mathbf{y}).
\end{align*}
Define 
\begin{align*}
    \mathcal{E}_n \coloneqq (\overline{P} \circ W)\left(\frac{1}{n} \log \frac{W(\mathbf{Y}|\mathbf{X})}{\overline{P}W(\mathbf{Y})} \leq C(\Gamma) + \frac{r}{\sqrt{n}} \right).
\end{align*}

\textbf{Analysis:}
We first write 
\begin{align}
	\mathcal{E}_n &= \sum_{j=1}^{k + 2} p_j \mathbb{P}_{\mathbf{X} \sim \text{Unif}(S^{n-1}_{R_j})} \left(\frac{1}{n} \log \frac{W(\mathbf{Y}|\mathbf{X})}{\overline{P}W(\mathbf{Y})} \leq C(\Gamma) + \frac{r}{\sqrt{n}} \right).   \label{epsnu} 
\end{align}
To proceed further, we upper bound 
\begin{align}
    &\mathbb{P}_{\mathbf{X} \sim \text{Unif}(S^{n-1}_{R_j})} \left(\frac{1}{n} \log \frac{W(\mathbf{Y}|\mathbf{X})}{\overline{P}W(\mathbf{Y})} \leq C(\Gamma) + \frac{r}{\sqrt{n}}   \right) \label{referbackn}\\
    &= \int_{\mathbf{y}  \in \mathbb{R}^n}  d \mathbf{y} Q_j^{cc}(\mathbf{y}) \mathbb{P} \left(\frac{1}{n} \log \frac{W(\mathbf{y}|\mathbf{X})}{\overline{P}W(\mathbf{y})} \leq C(\Gamma) + \frac{r}{\sqrt{n}} \Big | \mathbf{Y} = \mathbf{y} \right) \notag \\
    &\leq \int_{\mathbf{y} \in \mathbb{R}^n } d\mathbf{y} Q_j^{cc}(\mathbf{y}) \mathbb{P} \left(\frac{1}{n} \log \frac{W(\mathbf{y}|\mathbf{X})}{Q_i^{cc}(\mathbf{y})} \leq C(\Gamma) + \frac{r}{\sqrt{n}} \Big |  \mathbf{Y} = \mathbf{y} \right), \label{5cp} 
\end{align}
where $i \in \{1,2,\ldots, k + 2 \}$ depends on $\mathbf{y}$ and is such that $Q_i^{cc}(\mathbf{y})$ assigns the highest probability to $\mathbf{y}$. Continuing,
\begin{align}
&\int_{\mathbf{y} \in \mathbb{R}^n } d\mathbf{y} Q_j^{cc}(\mathbf{y}) \mathbb{P} \left( \log \frac{W(\mathbf{y}|\mathbf{X})}{Q_i^{cc}(\mathbf{y})} \leq n C(\Gamma) + r\sqrt{n} \Big |  \mathbf{Y} = \mathbf{y} \right) \notag \\
    &= \int_{\mathbf{y} \in \mathbb{R}^n } d\mathbf{y} Q_j^{cc}(\mathbf{y}) \mathbb{P} \left( \log \frac{W(\mathbf{y}|\mathbf{X})}{Q_j^{*}(\mathbf{y})} \leq nC(\Gamma) + r \sqrt{n} + \log \frac{Q_i^{cc}(\mathbf{y})}{Q_j^*(\mathbf{y})}  \Big |  \mathbf{Y} = \mathbf{y} \right)\\
    &\leq \int_{\mathbf{y} \in \mathbb{R}^n } d\mathbf{y} Q_j^{cc}(\mathbf{y}) \mathbb{P} \left( \log \frac{W(\mathbf{y}|\mathbf{X})}{Q_j^{*}(\mathbf{y})} \leq nC(\Gamma) + r \sqrt{n} + \kappa  \Big |  \mathbf{Y} = \mathbf{y} \right) + \delta_n^{(j)} \label{2x}
\end{align}
for sufficiently large $n$. In the last inequality above, we used Lemma \ref{slightlybetterlemma}. Specifically, in Lemma \ref{slightlybetterlemma}, let 
\begin{itemize}
\item $\Gamma = \Gamma_i$,
\item  $\Gamma' = \Gamma_j$,
    \item  $\epsilon = \Gamma_j - \Gamma_i$ so that $\epsilon = O\left(\frac{1}{\sqrt{n}} \right)$, and 
    \item $\Delta = \frac{\Gamma + N - |\epsilon|}{2}$.
\end{itemize}
Consequently, $\kappa$ is a constant from the result of Lemma \ref{slightlybetterlemma} and 
\begin{align*}
    \delta_n^{(j)} = Q_j^{cc} \left( \Bigg | \frac{||\mathbf{Y}||^2}{n} - \Gamma_j - N \Bigg | > \Delta \right).
\end{align*}
It can be verified that for $\mathbf{Y} \sim Q_j^{cc}$, $\mathbb{E}\left [ ||\mathbf{Y}||^2 \right] = n \Gamma_j + n N$ and $\text{Var}(||\mathbf{Y}||^2) = 4 n N \Gamma_j  + 2n N^2$. Thus, we have $\delta_n^{(j)} \to 0$ as $n \to \infty$ using Chebyshev inequality.

Continuing the derivation from $(\ref{2x})$, we have 
\begin{align}
    &\int_{\mathbf{y} \in \mathbb{R}^n } d\mathbf{y} Q_j^{cc}(\mathbf{y}) \mathbb{P} \left( \log \frac{W(\mathbf{y}|\mathbf{X})}{Q_j^{*}(\mathbf{y})} \leq nC(\Gamma) + r \sqrt{n} + \kappa   \Big |  \mathbf{Y} = \mathbf{y} \right) + \delta_n^{(j)} \notag \\
    &=  \mathbb{P}_{\mathbf{X} \sim \text{Unif}(S^{n-1}_{R_j})} \left( \log \frac{W(\mathbf{Y}|\mathbf{X})}{Q_j^{*}(\mathbf{Y})} \leq nC(\Gamma) + r \sqrt{n} + \kappa  \right) + \delta_n^{(j)} \notag \\
    &= \mathbb{P}_{\mathbf{X} \sim \text{Unif}(S^{n-1}_{R_j})} \left( \sum_{m=1}^{n} \log \frac{W(Y_m|X_m)}{Q_j^{*}(Y_m)} - n C(\Gamma_j) \leq n\left( C(\Gamma) - C(\Gamma_j) \right) + r \sqrt{n} + \kappa   \right) + \delta_n^{(j)} \notag \\
    &\stackrel{(a)}{=} \mathbb{P}_{\mathbf{X} \sim \text{Unif}(S^{n-1}_{R_j})} \left( \sum_{m=1}^{n}\left [ \log \frac{W(Y_m|X_m)}{Q_j^{*}(Y_m)} - \mathbb{E}\left[ \log \frac{W(Y_m|X_m)}{Q_j^{*}(Y_m)}\right] \right] \leq n\left( C(\Gamma) - C(\Gamma_j) \right) + r \sqrt{n} + \kappa \right) + \delta_n^{(j)} \notag \\
    &\stackrel{(b)}{=} \mathbb{P} \left( \sum_{m=1}^{n} \tilde{T}_m \leq n\left( C(\Gamma) - C(\Gamma_j) \right) + r \sqrt{n} + \kappa  \right) + \delta_n^{(j)} \notag \\
    &\stackrel{(c)}{=} \mathbb{P} \left( \frac{1}{\sqrt{n V(\Gamma_j)}} \sum_{m=1}^{n} \tilde{T}_m \leq \sqrt{n} \frac{ C(\Gamma) - C(\Gamma_j) }{\sqrt{V(\Gamma_j)}} + \frac{r}{\sqrt{V(\Gamma_j)}} + \frac{\kappa}{\sqrt{n V(\Gamma_j)}}  \right) + \delta_n^{(j)}  \notag \\
    &\stackrel{(d)}{\leq} \Phi\left( \sqrt{n} \frac{ C(\Gamma) - C(\Gamma_j) }{\sqrt{V(\Gamma_j)}} + \frac{r}{\sqrt{V(\Gamma_j)}} + \frac{\kappa }{\sqrt{n V(\Gamma_j)}}\right) + \delta_n \notag \\
    &\stackrel{(e)}{\leq} \Phi\left( \frac{\sqrt{n}}{2 \sqrt{V(\Gamma_j)}}\left(\frac{\Gamma - \Gamma_j}{N + \Gamma} \right) + \frac{r}{\sqrt{V(\Gamma_j)}} + \frac{\kappa'}{\sqrt{n V(\Gamma_j)}}\right) + \delta_n. \label{subsintosumof3}
\end{align}

In equality $(a)$ above, we used Lemma \ref{oftusedlemma}. Specifically, use Equation $(\ref{equsentimes})$ in Lemma \ref{oftusedlemma} with $\Gamma = \Gamma_j$. 

In equality $(b)$ above, we used~\cite[Lemma
2]{mahmood2025channelcodinggaussianchannels},
where the $\tilde{T}_m$'s are i.i.d. and 
\begin{align}
    \tilde{T}_m = \log \frac{W(Y|\sqrt{\Gamma_j})}{Q^*_j(Y)} - \mathbb{E} \left [\log  \frac{W(Y|\sqrt{\Gamma_j})}{Q^*_j(Y)} \right ],     
\end{align}
where $Y \sim \mathcal{N}(\sqrt{\Gamma_j}, N)$. 

In equality $(c)$ above, we normalize the sum to have unit variance, which follows from Lemma~\ref{oftusedlemma}.

In inequality $(d)$ above, we used the Berry-Esseen Theorem \cite{esseen11} to obtain convergence of the CDF of the normalized sum of i.i.d. random variables $\tilde{T}_m$'s to the standard normal CDF, with $\delta_n \to 0$ accounting for both the rate of convergence and $\delta_n^{(j)} \to 0$. In inequality $(e)$ above, we used a Taylor series approximation $C(\Gamma) = C(\Gamma_j) + C(\tilde{\Gamma})(\Gamma - \Gamma_j)$ for some $\tilde{\Gamma}$ between $\Gamma$ and $\Gamma_j$. Then we further used the fact that $\Gamma - \Gamma_j = O(1/\sqrt{n})$ so that there exists a constant $\kappa'$ for which inequality $(e)$ holds. 

We can now upper bound $(\ref{referbackn})$ by $(\ref{subsintosumof3})$, which allows us to upper bound $(\ref{epsnu})$ as 
\begin{align*}
    \mathcal{E}_n &\leq \sum_{j=1}^{k+2} p_j \Phi\left( \frac{\sqrt{n}}{2 \sqrt{V(\Gamma_j)}}\left(\frac{\Gamma - \Gamma_j}{N + \Gamma} \right) + \frac{r}{\sqrt{V(\Gamma_j)}} + \frac{\kappa'}{\sqrt{n V(\Gamma_j)}}\right) + \delta_n
\end{align*}
for some redefined sequence $\delta_n \to 0$ as $n \to \infty$. Using Equation $(\ref{38n})$ and the formula for $C'(\Gamma)$ from Lemma \ref{oftusedlemma}, we can simplify the upper bound as 
\begin{align*}
     \mathcal{E}_n &\leq \sum_{j=1}^{k+2} p_j \Phi\left( -\frac{C'(\Gamma)}{ \sqrt{V(\Gamma_j)}}u_j + \frac{r}{\sqrt{V(\Gamma_j)}} + \frac{\kappa'}{\sqrt{n V(\Gamma_j)}}\right) + \delta_n. 
\end{align*}
Therefore, since $\Gamma_j \to \Gamma$ as $n \to \infty$, we have 
\begin{align}
    \limsup_{n \to \infty} \mathcal{E}_n &\leq \sum_{j=1}^{k+2} p_j \Phi\left(\frac{r}{\sqrt{V(\Gamma)}} - \frac{C'(\Gamma)}{\sqrt{V(\Gamma)}} u_j \right) \notag \\
    &= \mathbb{E}_{P_U} \left [  \Phi\left( \frac{r}{\sqrt{V(\Gamma)}} - \frac{C'(\Gamma)U}{\sqrt{V(\Gamma)}}  \right) \right ]. \notag  
\end{align} 
Since $P_U \in \mathcal{U}_{f, \mathbf{\Gamma}}^{(k)}$ was arbitrary, we obtain 
\begin{align}
    \limsup_{n \to \infty} \mathcal{E}_n \leq \inf_{P_U \in \mathcal{U}_{f, \mathbf{\Gamma}}^{(k)}} \mathbb{E}_{P_U} \left [  \Phi\left( \frac{r}{\sqrt{V(\Gamma)}} - \frac{C'(\Gamma)U}{\sqrt{V(\Gamma)}}  \right) \right ].   \label{extendtoerror}
\end{align}
To finish the proof, we state the achievability result in terms of an upper bound on the minimum average probability of error $\epsilon$ of channel codes in $\mathcal{C}_{n, R, f, \mathbf{\Gamma}}$ $(n, R, \Gamma, V)$ for a rate $R = C(\Gamma) + \frac{r}{\sqrt{n}}$. From Lemma \ref{aaron'slemma}, we have for $\theta = 1/n^{\vartheta}$ for $1 > \vartheta > 1/2$,
\begin{align}
    \epsilon \leq (\overline{P} \circ W) \left(\frac{1}{n} \log \frac{W(\mathbf{Y}|\mathbf{X})}{\overline{P}W(\mathbf{Y})} \leq C(\Gamma) + \frac{r}{\sqrt{n}} + \frac{1}{n^\vartheta} \right) + e^{-n^{1-\vartheta}}.  \label{notplaying}
\end{align}
For any $r' > r$, we have $\frac{r}{\sqrt{n}} + \frac{1}{n^\vartheta} < \frac{r'}{\sqrt{n}}$ eventually, so
\begin{align}
    \limsup_{n \to \infty} \epsilon &\leq \limsup_{n \to \infty} (\overline{P} \circ W) \left(\frac{1}{n} \log \frac{W(\mathbf{Y}|\mathbf{X})}{\overline{P}W(\mathbf{Y})} \leq C(\Gamma) + \frac{r'}{\sqrt{n}} \right) \notag\\
    &\leq \inf_{P_U \in \mathcal{U}_{f, \mathbf{\Gamma}}^{(k)}} \mathbb{E}_{P_U} \left [  \Phi\left( \frac{r'}{\sqrt{V(\Gamma)}} - \frac{C'(\Gamma)U}{\sqrt{V(\Gamma)}}  \right) \right ], \label{notplayggg}
\end{align}
where the last inequality above follows from $(\ref{extendtoerror})$. From Lemma \ref{contrlemma}, we have that the RHS of $(\ref{notplayggg})$ is continuous in the variable $r'$. Hence, letting $r' \to r$ in $(\ref{notplayggg})$ establishes the result.

\section{Concluding Remarks}
\label{sec:conclusion}

We examined channel coding performance under more refined cost constraints than have been considered previously. The proposed framework unifies and extends several prior works. Although it was not our focus, it should be noted that, compared with an almost sure cost constraint, the cost constraints considered here generally provide for improved coding performance \cite{mahmood2024improvedchannelcodingperformance}, \cite{mahmood2025channelcodinggaussianchannels}.

We have focused on channel coding in the normal approximation regime.
As noted earlier, specializations of our framework have been considered within the realm of lossy source coding. 
It would be interesting to lift those results into the general framework proposed here. Another natural extension would be to prove a version of Theorem~\ref{main_thm} for discrete memoryless channels. It would also be of interest to extend the framework to the moderation deviations~\cite{Altug:MDP,Altug:ISIT10} and error exponent regimes.

\appendices 

\section{Excess Cost Probability Constraint}
\label{app:excess}

We have, for any $U$ satisfying $\mathbb{E}[U] \le 0$ and
$\mathbb{E}[f_1^{(\alpha)}(U)] \le \delta$,
\begin{align}
    \mathbb{E} \left [ \Phi\left( \frac{r}{\sqrt{V(\Gamma)}} - \frac{C'(\Gamma)U}{\sqrt{V(\Gamma)}}  \right) \right ] 
     \ge 
    \Phi\left( \frac{r}{\sqrt{V(\Gamma)}} - \frac{C'(\Gamma)\gamma}{\sqrt{V(\Gamma)}}  \right)
    P(U \le \gamma).
\end{align}
But
\begin{equation}
    \delta \ge \mathbb{E}[f_1^{(\alpha)}(U)] \ge P(U > \gamma).
\end{equation}
On the other hand, for any $\alpha > 0$, we can choose
\begin{equation}
    P(u_i) = \pi_i \quad i = 1,2,3,
\end{equation}
where, e.g.,
\begin{align}
    \pi_1 & = \frac{\delta}{1 + \sqrt{\alpha}} \\
    \pi_2 & =  1 - \delta - \alpha \\
    \pi_3 & = 1 - \pi_1 - \pi_2 \\
    u_1 & = \gamma + \frac{1}{\sqrt{\alpha}} \\
    u_2 & = \gamma \\
    u_3 & = - \frac{u_1 \pi_1 + u_2 \pi_2}{\pi_3}.
\end{align}
One can verify that for this choice,
\begin{align}
    \lim_{\alpha \rightarrow 0}
     \mathbb{E} \left [ \Phi\left( \frac{r}{\sqrt{V(\Gamma)}} - \frac{C'(\Gamma)U}{\sqrt{V(\Gamma)}}  \right) \right ] 
     = 
     (1-\delta) \Phi\left( \frac{r}{\sqrt{V(\Gamma)}} - \frac{C'(\Gamma)\gamma}{\sqrt{V(\Gamma)}}  \right).
\end{align}

\section{Proof of Lemma \ref{contrlemma} \label{contrlemma_proof}}

Recall that the function $g(r)$ is given by 
\begin{align}
    g(r) = \inf_{\substack{P_{U} \in \mathcal{U}_{f, \mathbf{\Gamma}}\\ |\text{supp}(P_U)| \leq k + 2 }} \mathbb{E}_{P_U} \left [  \Phi\left( \kappa_1 \,r - \kappa_2 \,U  \right) \right ]. \label{bxe} 
\end{align}
Fix any $r_1, r_2$ such that $r_1 \neq r_2$. Fix any small $\epsilon > 0$. Choose $P_{\epsilon, 1}$ in the feasible set of $(\ref{bxe})$ such that\\ $\mathbb{E}_{P_{\epsilon, 1}}\left [\Phi(\kappa_1r_1 - \kappa_2U) \right] \leq g(r_1) + \epsilon$. We then have 
\begin{align*}
    &g(r_1) - \mathbb{E}_{P_{\epsilon, 1}}\left [\Phi(\kappa_1r_2 - \kappa_2 U) \right]\\
    &\geq \mathbb{E}_{P_{\epsilon, 1}}\left [\Phi(\kappa_1r_1 - \kappa_2U) \right] - \mathbb{E}_{P_{\epsilon, 1}}\left [\Phi(\kappa_1r_2 - \kappa_2U) \right] - \epsilon\\
    &= \mathbb{E}_{P_{\epsilon, 1}}\left [ \Phi(\kappa_1r_1 - \kappa_2U) - \Phi(\kappa_1r_2 -\kappa_2 U)   \right ] - \epsilon.
\end{align*}
Since $\Phi$ is Lipschitz continuous with Lipschitz constant $\frac{1}{\sqrt{2\pi}}$, we have 
\begin{align}
     g(r_1) - \mathbb{E}_{P_{\epsilon, 1}}\left [\Phi(\kappa_1r_2 - \kappa_2U) \right]  \geq - \frac{\kappa_1}{\sqrt{2\pi}} |r_1 - r_2| - \epsilon.  \label{gp31}
\end{align}
Note also that 
\begin{align}
    g(r_2) \leq \mathbb{E}_{P_{\epsilon, 1}}\left [\Phi(\kappa_1r_2 - \kappa_2U) \right]. \label{gp32}
\end{align}
From $(\ref{gp31})$ and $(\ref{gp32})$, we obtain 
\begin{align}
   g(r_1) - g(r_2) &\geq - \frac{\kappa_1}{\sqrt{2\pi}} |r_1 - r_2|- \epsilon. \label{b2vv} 
\end{align}
The conclusion follows by swapping the roles of $r_1$ and $r_2$ and
letting $\epsilon \rightarrow 0$.

\section{Proof of Lemma \ref{log_dist_eq} \label{log_dist_eq_proof}}

We have
\begin{align*}
    W(\mathbf{y}|\mathbf{x}) & = \frac{1}{(2 \pi N)^{n/2}} \exp \left( - \frac{||\mathbf{y} - \mathbf{x}||^2}{2N} \right) \\
    q(\mathbf{y}) & = \frac{1}{(2\pi (\Gamma + N))^{n/2}} \exp \left( -\frac{||\mathbf{y}||^2}{2(\Gamma + N )} \right).
\end{align*}

Since $(\mathbf{X}, \mathbf{Y}) \sim \overline{P} \circ W$, we have $\mathbf{Y} = \mathbf{X} + \mathbf{Z}$, where $\mathbf{X}$ and $\mathbf{Z}$ are independent and $\mathbf{Z} \sim \mathcal{N}(\mathbf{0}, N \mathbf{I}_n)$. Then   

\begin{align*}
    \log \frac{W(\mathbf{Y}|\mathbf{X})}{q(\mathbf{Y})} &= \frac{n}{2} \log \left(1 + \frac{\Gamma}{N} \right)  - \frac{||\mathbf{Y} - \mathbf{X}||^2}{2N} +  \frac{||\mathbf{Y}||^2}{2(N + \Gamma )} \notag \\
    &= n C(\Gamma) -  \frac{||\mathbf{Z}||^2}{2N} +  \frac{||\mathbf{X} + \mathbf{Z}||^2}{2(N + \Gamma )} \notag  \\
    &= n C(\Gamma)  + \sum_{i=1}^n \frac{(X_i + Z_i)^2}{2(N + \Gamma )} -\frac{1}{2N} Z_i^2 \notag \\
    &= n C(\Gamma)  + \sum_{i=1}^n \frac{X_i^2}{2\Gamma} - \sum_{i=1}^n \frac{\Gamma}{2N(N + \Gamma)} \left( Z_i - \frac{NX_i}{\Gamma} \right)^2 \notag \\
    &= n C(\Gamma)  + \frac{nS}{2\Gamma} - \sum_{i=1}^n \frac{\Gamma}{2N(N + \Gamma)} \left( Z_i - \frac{NX_i}{\Gamma} \right)^2 \notag \\
    &\stackrel{d}{=} n C(\Gamma)  + \frac{nS}{2\Gamma} - \sum_{i=1}^n \frac{\Gamma}{2N(N + \Gamma)} \left( Z_i - \frac{N \sqrt{S}}{\Gamma} \right)^2 \notag \\ 
    &\stackrel{d}{=} n C(\Gamma)  + \frac{nS}{2\Gamma} - \sum_{i=1}^n \frac{\Gamma}{2(N + \Gamma)} \left( \tilde{Z}_i - \frac{\sqrt{N S}}{\Gamma}  \right)^2.
\end{align*}
In the penultimate equality above, we used spherical symmetry w.r.t. $\mathbf{X}$. In the last equality, we substituted $\mathbf{Z} = \sqrt{N} \tilde{\mathbf{Z}}$.

\section{Proof of Lemma \ref{lemma_gen_lower_bnd_wo_opt} \label{lemma_gen_lower_bnd_wo_opt_proof}}

For $\mathbf{X} \sim \overline{P}$, let $S \sim P_{S}$. Using Lemma \ref{log_dist_eq}, we first write  
\begin{align}
    &(\overline{P} \circ W)\left( \log \frac{W(\mathbf{Y}|\mathbf{X})}{q(\mathbf{Y})} \leq n \gamma  \right) \notag \\
    &=  (\overline{P} \circ W)\left(n C(\Gamma)  + \frac{nS}{2\Gamma} - \sum_{i=1}^n \frac{\Gamma}{2(N + \Gamma)} \left( \tilde{Z}_i - \frac{\sqrt{N S}}{\Gamma}  \right)^2 \leq n \gamma  \right) \notag \\
    &= \int_0^\infty d P_{S}(s) \mathbb{P}\left(n C(\Gamma)  + \frac{ns}{2\Gamma} - \sum_{i=1}^n \frac{\Gamma}{2(N + \Gamma)} \left( \tilde{Z}_i - \frac{\sqrt{N s}}{\Gamma}  \right)^2 \leq n \gamma  \right).  \label{f2b}
\end{align}

To proceed further, we lower bound 
\begin{align*}
    \mathbb{P}\left(n C(\Gamma)  + \frac{ns}{2\Gamma} - \sum_{i=1}^n \frac{\Gamma}{2(N + \Gamma)} \left( \tilde{Z}_i - \frac{\sqrt{N s}}{\Gamma}  \right)^2 \leq n \gamma  \right).    
\end{align*}
We have 
\begin{align*}
    &\mathbb{P}\left(n C(\Gamma)  + \frac{ns}{2\Gamma} - \sum_{i=1}^n \frac{\Gamma}{2(N + \Gamma)} \left( \tilde{Z}_i - \frac{\sqrt{N s}}{\Gamma}  \right)^2 \leq n \gamma  \right)\\
    &=  \mathbb{P}\left(  \frac{\Gamma}{2(N + \Gamma)} \sum_{i=1}^n  \left( \tilde{Z}_i - \frac{\sqrt{N s}}{\Gamma}  \right)^2 \geq n \left( \frac{s}{2\Gamma} - \gamma \right)  + nC(\Gamma)  \right)
\end{align*}

Note that
$$D_n := \sum_{i=1}^n  \left( \tilde{Z}_i - \frac{\sqrt{N s}}{\Gamma}  \right)^2 \sim \chi^2_n\left( \frac{nNs}{ \Gamma^2} \right)$$
so that   
\begin{align*}
    \mathbb{E}[D_n] &= n +  \frac{nNs}{ \Gamma^2} = \frac{n (\Gamma^2 + Ns)}{\Gamma^2},\\
    \text{Var}(D_n) &= 2\left(n + \frac{2n N s}{\Gamma^2}  \right) = 2n \frac{\Gamma^2 + 2Ns}{ \Gamma^2}.
\end{align*}
Define 
\begin{align*}
    D_{n, i} := \left( \tilde{Z}_i - \frac{\sqrt{N s}}{\Gamma}  \right)^2. 
\end{align*}
Then 
\begin{align*}
    &\mathbb{P}\left(  \frac{\Gamma}{2(N + \Gamma)} \sum_{i=1}^n  D_{n, i} \geq n \left( \frac{s}{2\Gamma} - \gamma \right)  + nC(\Gamma)  \right)\\
    &= \mathbb{P}\left( \frac{1}{\sqrt{\operatorname{Var}(D_n)}} \sum_{i=1}^n  \left [ D_{n, i} - \left(1 + \frac{Ns}{\Gamma^2} \right) \right] \geq \frac{\sqrt{2n}}{\sqrt{\Gamma^2 + 2Ns}} \left( \frac{s}{2\Gamma} - \gamma \right)(N + \Gamma)  + \frac{\sqrt{2n}C(\Gamma)(N + \Gamma)}{ \sqrt{\Gamma^2 + 2 N s}} - \mbox{} \right. \\
    & \left . \quad \quad \quad \quad \quad \quad \quad \quad \frac{\sqrt{n}\Gamma}{\sqrt{2} \sqrt{\Gamma^2 + 2 N s}} - \frac{\sqrt{n} N s}{\Gamma\sqrt{2} \sqrt{\Gamma^2 + 2Ns}} \right)\\
    &= \mathbb{P}\left( \frac{1}{\sqrt{\operatorname{Var}(D_n)}} \sum_{i=1}^n  \overline{D}_{n, i} \geq \frac{\sqrt{2n}(N + \Gamma)}{\sqrt{\Gamma^2 + 2 N s}}\left [\frac{s}{2\Gamma} - \gamma + C(\Gamma) \right] - \frac{\sqrt{n}(\Gamma^2 + Ns)}{\Gamma \sqrt{2} \sqrt{\Gamma^2 + 2Ns}} \right)\\
    &= \mathbb{P}\left( \frac{1}{\sqrt{\operatorname{Var}(D_n)}} \sum_{i=1}^n  \overline{D}_{n, i} \geq \frac{\sqrt{2n}(N + \Gamma)(C(\Gamma) - \gamma)}{\sqrt{\Gamma^2 + 2 N s}} + \frac{\sqrt{n}(s - \Gamma)}{\sqrt{2}\sqrt{\Gamma^2 + 2 N s}}  \right),
\end{align*}
where 
\begin{align*}
    D_{n, i} &\sim \chi^2_1\left (\frac{Ns}{\Gamma^2} \right)\\
    \overline{D}_{n, i} &= D_{n, i} - \left(1 + \frac{Ns}{\Gamma^2} \right)\\
    \mathbb{E}[\overline{D}_{n, i}] &= 0\\
    \sigma_i^2 := \mathbb{E}[\overline{D}_{n, i}^2] &= 2 \left(1 + \frac{2Ns}{\Gamma^2} \right)
    \\
    \rho_i := \mathbb{E}\left [ |\overline{D}_{n, i}|^3 \right] 
                                                    &\leq \left(\mathbb{E}\left [ \left ( D_{n, i} - \left(1 + \frac{Ns}{\Gamma^2}\right) \right )^4 \right] \right)^{3/4}\\
    &= \left( 12\left(1 + 2 \frac{Ns}{\Gamma^2}  \right)^2 + 48 \left(1 + \frac{4Ns}{\Gamma^2} \right) \right)^{3/4}\\
    &= \left(60 + 48 \frac{N^2s^2}{\Gamma^4} + 240 \frac{N s}{\Gamma^2} \right)^{3/4}.
\end{align*}
Hence, by the Berry-Esseen Theorem~(e.g.,~\cite{esseen11}), 
\begin{align}
    &\mathbb{P}\left( \frac{1}{\sqrt{\operatorname{Var}(D_n)}} \sum_{i=1}^n  \overline{D}_{n, i} \geq \frac{\sqrt{2n}(N + \Gamma)(C(\Gamma) - \gamma)}{\sqrt{\Gamma^2 + 2 N s}} + \frac{\sqrt{n}(s - \Gamma)}{\sqrt{2}\sqrt{\Gamma^2 + 2 N s}}  \right)\\
    &\geq 1 - \Phi\left( \frac{\sqrt{2n}(N + \Gamma)(C(\Gamma) - \gamma)}{\sqrt{\Gamma^2 + 2 N s}} + \frac{\sqrt{n}(s - \Gamma)}{\sqrt{2}\sqrt{\Gamma^2 + 2 N s}} \right) - \frac{n \left(60 + 48 \frac{N^2s^2}{\Gamma^4} + 240 \frac{N s}{\Gamma^2} \right)^{3/4}}{\left(2n \left(1 + \frac{2Ns}{\Gamma^2} \right) \right)^{3/2}}\\
    &= 1 - \Phi\left( \frac{\sqrt{2n}(N + \Gamma)(C(\Gamma) - \gamma)}{\sqrt{\Gamma^2 + 2 N s}} + \frac{\sqrt{n}(s - \Gamma)}{\sqrt{2}\sqrt{\Gamma^2 + 2 N s}} \right) - \frac{1}{\sqrt{n}} \left(15 - \frac{48N^2 s^2}{\Gamma^4 + 4 N s \Gamma^2 + 4N^2 s^2} \right)^{3/4}\\
    &\geq 1 - \Phi\left( \frac{\sqrt{2n}(N + \Gamma)(C(\Gamma) - \gamma)}{\sqrt{\Gamma^2 + 2 N s}} + \frac{\sqrt{n}(s - \Gamma)}{\sqrt{2}\sqrt{\Gamma^2 + 2 N s}} \right) - \frac{15^{3/4}}{\sqrt{n}}. \label{e2b}
\end{align}
Substituting $(\ref{e2b})$ in $(\ref{f2b})$, we obtain 
\begin{align*}
    &(\overline{P} \circ W)\left( \log \frac{W(\mathbf{Y}|\mathbf{X})}{q(\mathbf{Y})} \leq n \gamma  \right)\\
    &\geq 1 - \frac{15^{\frac{3}{4}}}{\sqrt{n}} -  \mathbb{E}_{P_S} \left [ \Phi\left( \frac{\sqrt{2n}(N + \Gamma)(C(\Gamma) - \gamma)}{\sqrt{\Gamma^2 + 2 N S}} + \frac{\sqrt{n}(S - \Gamma)}{\sqrt{2}\sqrt{\Gamma^2 + 2 N S}}  \right)  \right].
\end{align*}

\section{Proof of Lemma \ref{sncompact} \label{sncompactproof}}

We first show that $\mathcal{S}_{n, f, \mathbf{\Gamma}}$ is tight~\cite[p.~59]{Billingsley:Convergence2}. Note that for every $\epsilon > 0$, there exists a compact set $K_\epsilon = \left [0, \frac{\Gamma}{\epsilon} \right]$ such that 
\begin{align*}
    \inf_{P_S \in \mathcal{S}_{n, f, \mathbf{\Gamma}}} \, P_S\left(K_\epsilon \right) \geq 1 - \epsilon. 
\end{align*}
This follows from Markov's inequality: 
\begin{align*}
    \mathbb{P}_{P_S}\left(S > \frac{\Gamma}{\epsilon} \right) \leq \frac{\mathbb{E}[S]}{\Gamma/\epsilon} \leq \epsilon
\end{align*}
for every $P_S \in \mathcal{S}_{n, f, \mathbf{\Gamma}}$. 

Hence, the set $\mathcal{S}_{n, f, \mathbf{\Gamma}}$ is tight. Therefore, by Prokhorov's theorem~\cite[Thm.~5.1]{Billingsley:Convergence2}, every sequence of probability measures in $\mathcal{S}_{n, f, \mathbf{\Gamma}}$ has a weakly convergent subsequence that converges to some probability measure in the closure of $\mathcal{S}_{n, f, \mathbf{\Gamma}}$. We next show that $\mathcal{S}_{n, f, \mathbf{\Gamma}}$ is closed in the weak topology. Consider a sequence $P_{S_k} \in \mathcal{S}_{n, f, \mathbf{\Gamma}}$ that converges weakly to some $P_S$, i.e.,
\begin{align*}
    \int \varphi d P_{S_k} \to \int \varphi d P_S \quad \text{ as } \quad  k \to \infty  
\end{align*}
for every bounded, continuous function $\varphi$. Let $\mathbb{E}_k[\cdot]$ and $\mathbb{E}[\cdot]$ denote expectations w.r.t. $P_{S_k}$ and $P_S$, respectively. If we define a function $g(x) = \max\{0, x\}$, then by the Portmanteau theorem for lower semicontinuous functions bounded from below~\cite[Thm.~4.4.4]{Chung:Probability3},
\begin{align*}
    \mathbb{E}[S] \leq \mathbb{E}[g(S)] \leq \liminf_k \mathbb{E}_k[g(S)] \leq \sup_k \mathbb{E}_k [g(S)] = \sup_k \mathbb{E}_k [S] \leq \Gamma,  
\end{align*}
where the equality above follows from the fact that $\text{supp}(P_{S_k}) \subset [0, \infty)$ for every $k$ by definition of $\mathcal{S}_{n ,f, \mathbf{\Gamma}}$. Furthermore, by the Portmanteau theorem, since each $f_i$ is nonnegative and lower semicontinuous,  
\begin{align*}
    \mathbb{E}[f_i(\sqrt{n}(S - \Gamma))] \leq \liminf_k \mathbb{E}_k[f_i(\sqrt{n}(S - \Gamma))] \leq \sup_k \mathbb{E}_k[f_i(\sqrt{n}(S - \Gamma))] \leq \Gamma_i.  
\end{align*}
Lastly, we also show that $\text{supp}(P_S) \subset [0, \infty)$. This again follows from the Portmanteau theorem since for the closed set $C = [0, \infty)$, we have $P_{S}(C) \geq \limsup_k P_{S_k}(C) = 1$.
It follows that every sequence in $\mathcal{S}_{n, f, \mathbf{\Gamma}}$ 
has a subsequence that converges weakly to a distribution in $\mathcal{S}_{n, f, \mathbf{\Gamma}}$; hence, $\mathcal{S}_{n, f, \mathbf{\Gamma}}$ is sequentially compact w.r.t. the topology of weak convergence. But weak convergence in $\mathcal{P}(\mathbb{R})$ is equivalent to convergence in
the Prokhorov metric~\cite[Thm.~6.8]{Billingsley:Convergence2} and thus 
$\mathcal{S}_{n, f,     \mathbf{\Gamma}}$ is sequentially compact under
the Prokhorov metric and hence compact.

\section{Proof of Lemma \ref{lemminac} \label{lemminacproof}}

Define a function $\phi_{n, \gamma} : [0, \infty) \to [0, 1]$ as 
\begin{align*}
    \phi_{n, \gamma}(s) &=   \Phi\left( \frac{\sqrt{2n}(N + \Gamma)( \gamma - C(\Gamma))}{\sqrt{\Gamma^2 + 2 N s}} + \frac{\sqrt{n}(\Gamma - s)}{\sqrt{2}\sqrt{\Gamma^2 + 2 N s}}  \right).
\end{align*}

Define the continuous, linear functional $\mathcal{L}(P) = \mathbb{E}_P\left [ \phi_{n, \gamma}(S) \right]$. Consider  
\begin{align}
    \inf_{P_S \in \mathcal{S}_{n, f, \mathbf{\Gamma}}} \mathcal{L}(P_S). \label{4b}
\end{align}

Now consider a sequence $P_{S_k} \in \mathcal{S}_{n, f, \mathbf{\Gamma}}$ such that 
\begin{align*}
    \mathbb{E}_k[\phi_{n, \gamma}(S)] \to \inf_{P_S \in \mathcal{S}_{n, f, \mathbf{\Gamma}} } \mathbb{E}_{P_S}[\phi_{n, \gamma}(S)].
\end{align*}
By compactness of $\mathcal{S}_{n, f, \mathbf{\Gamma}}$ (Lemma~\ref{sncompact}), there exists a subsequence $P_{S_m}$ that converges weakly to some $P_S^* \in \mathcal{S}_{n, f, \mathbf{\Gamma}}$. Since $\phi_{n,\gamma}(s)$ is a continuous, bounded function, we have that 
\begin{align*}
    \mathbb{E}_m[\phi_{n, \gamma}(S)] \to \mathbb{E}_{P_S^*}[\phi_{n, \gamma}(S)]
\end{align*}
by the Portmanteau theorem. Hence, 
\begin{align}
    \inf_{P_S \in \mathcal{S}_{n, f, \mathbf{\Gamma}}} \mathbb{E}_{P_S}[\phi_{n, \gamma}(S)] = \min_{P_S \in \mathcal{S}_{n, f, \mathbf{\Gamma}}} \mathbb{E}_{P_S}[\phi_{n, \gamma}(S)].  \label{nowamin}
\end{align}

\section{Proof of Lemma \ref{asymplemma} \label{asymplemmaproof}}

By letting $\gamma = C(\Gamma) + r/ \sqrt{n}$ in Corollary \ref{genlem3pt}, we obtain  
\begin{align}
    &\inf_{\overline{P} \in \mathcal{P}_{n, f, \mathbf{\Gamma}}}(\overline{P} \circ W) \left( \log  \frac{W(\mathbf{Y}|\mathbf{X})}{q(\mathbf{Y})} \leq n C(\Gamma) + r \sqrt{n} \right) \geq \notag \\
    &\quad \quad \quad \quad \quad \quad \quad \quad \quad \min_{ \substack{P_{S} \in \mathcal{S}_{n, f, \mathbf{\Gamma}}\\|\text{supp}(P_S)| \leq k+2} } \mathbb{E}_{P_S} \left [ \Phi\left( \frac{\sqrt{2}(N + \Gamma)r}{\sqrt{\Gamma^2 + 2 N S}} + \frac{\sqrt{n}(\Gamma - S)}{\sqrt{2}\sqrt{\Gamma^2 + 2 N S}}  \right)  \right] - \frac{15^{\frac{3}{4}}}{\sqrt{n}}. \label{b5my}
\end{align}
We first write
\begin{align}
    &\min_{ \substack{P_{S} \in \mathcal{S}_{n, f, \mathbf{\Gamma}}\\|\text{supp}(P_S)| \leq k+2} } \mathbb{E}_{P_S} \left [ \Phi\left( \frac{\sqrt{2}(N + \Gamma)r}{\sqrt{\Gamma^2 + 2 N S}} + \frac{\sqrt{n}(\Gamma - S)}{\sqrt{2}\sqrt{\Gamma^2 + 2 N S}}  \right)  \right]\\
    &= 1 - \max_{ \substack{P_{S} \in \mathcal{S}_{n, f, \mathbf{\Gamma}}\\|\text{supp}(P_S)| \leq k+2} } \mathbb{E}_{P_S} \left [ \Phi\left( -\frac{\sqrt{2}(N + \Gamma)r}{\sqrt{\Gamma^2 + 2 N S}} + \frac{\sqrt{n}(S - \Gamma)}{\sqrt{2}\sqrt{\Gamma^2 + 2 N S}}  \right)  \right]\\
    &= 1 - \max_{ \substack{P_{S} \in \mathcal{S}_{n, f, \mathbf{\Gamma}}\\|\text{supp}(P_S)| \leq k+2} } \mathbb{E}_{P_S} \left [ \psi_{n,r}(S) \right ], \label{fdnkgf}
\end{align}
where we define 
\begin{align*}
    \psi_{n, r}(s) &= \Phi\left( -\frac{\sqrt{2}(N + \Gamma)r}{\sqrt{\Gamma^2 + 2 N s}} + \frac{\sqrt{n}(s - \Gamma)}{\sqrt{2}\sqrt{\Gamma^2 + 2 N s}}  \right)\\
    &= \Phi\left( \frac{1}{\sqrt{2}\sqrt{\Gamma^2 + 2 N s}} \left [ \sqrt{n}(s - \Gamma) - 2(N + \Gamma) r  \right]  \right). 
\end{align*}
We now upper bound the second term in $(\ref{fdnkgf})$. Let $a_n = \frac{1}{n^{1/3}} $. Recall that we assume in Lemma \ref{asymplemma} that $f$ and $\mathbf{\Gamma}$ satisfy Condition \ref{reg_cond2} and hence,  enforce uniform upper-tail concentration as described in $(\ref{g13})$ and $(\ref{f1})$.

Define 
\begin{align*}
    \mathcal{P}_{n,1} &:= \left \{ s \in [0, \infty): 0 \leq s < \Gamma - a_n \right \},\\
    \mathcal{P}_{n,2} &:= \left \{s \in [0, \infty): \Gamma - a_n \leq s \leq \Gamma + a_n \right \},\\
    \mathcal{P}_{n,3} &:= \left \{ s \in [0, \infty): s > \Gamma + a_n \right \}.
\end{align*}
It is straightforward to check that for every $r \in \mathbb{R}$, since $a_n = \omega(n^{-1/2})$,
\begin{align*}
    \lim_{n \to \infty} \sup_{s \in \mathcal{P}_{n, 1}} \psi_{n, r}(s) = 0.
\end{align*}
This implies that $\sup_{s \in \mathcal{P}_{n, 1}}\psi_{n, r}(s) \leq \delta_n^{(1)}$ for some sequence $\delta_n^{(1)} \to 0$ as $n \to \infty$.

For $s \in \mathcal{P}_{n, 2}$, define $g(s) = \frac{1}{\sqrt{2}\sqrt{\Gamma^2 + 2Ns}}$ so that $g'(s) = - \frac{N}{\sqrt{2}(\Gamma^2 + 2 N s)^{3/2}}$. Since $g'(s)$ is continuous over the set $\mathcal{P}_{n,2}$, there exists a finite constant $L$ independent of $n$ such that  
\begin{align*}
    \sup_{s \in \mathcal{P}_{n,2}} |g'(s)| &\leq L.   
\end{align*}
Then we have
\begin{align*}
    &g(s) \sqrt{n}(s - \Gamma) - 2 g(s)(N + \Gamma) r - g(\Gamma) \sqrt{n}(s - \Gamma) + 2 g(\Gamma)(N + \Gamma)r\\
    &= \sqrt{n}(s - \Gamma) (g(s) - g(\Gamma)) + 2r(N + \Gamma) (g(\Gamma) - g(s)).  
\end{align*}
Furthermore, by mean value theorem, there exists a $\tilde{s} \in \mathcal{P}_{n,2}$ such that
\begin{align}
    &\big |\sqrt{n}(s - \Gamma) (g(s) - g(\Gamma)) + 2r(N + \Gamma) (g(\Gamma) - g(s)) \big |\\
    &\leq \sqrt{n} |s - \Gamma| \cdot |g(s) - g(\Gamma)| + 2 |r| (N + \Gamma) \cdot |g(\Gamma) - g(s)|\\
    &\leq \sqrt{n} |s - \Gamma|^2 \cdot |g'(\tilde{s})|  + 2 |r| (N + \Gamma) \cdot |s - \Gamma| \cdot |g'(\tilde{s})|\\
    &\leq \sqrt{n} L a_n^2    + 2L |r| (N + \Gamma) \cdot a_n. \label{rb}
\end{align}
This implies that 
\begin{align*}
    &\sup_{s \in \mathcal{P}_{n,2}}  \Bigg | \Phi\left( g(s) \left [ \sqrt{n}(s - \Gamma) - 2(N + \Gamma) r  \right]  \right) - \Phi\left( g(\Gamma) \left [ \sqrt{n}(s - \Gamma) - 2(N + \Gamma) r  \right]  \right)  \Bigg |\\
    &\leq \frac{L}{\sqrt{2\pi}}  \left(\sqrt{n}  a_n^2    + 2 |r| (N + \Gamma) \cdot a_n \right) =: \delta_n^{(2)},  
\end{align*}
where in the last inequality above, we used $(\ref{rb})$ and the fact that $\Phi$ is Lipschitz continuous with Lipschitz constant $\frac{1}{\sqrt{2\pi}}$. Note that $\delta_{n}^{(2)} \to 0$ as $n \to \infty$, since $a_n = o(n^{-1/4})$. 

For $\mathcal{P}_{n,3}$, we have 
\begin{align*}
    \sup_{P_S \in \mathcal{S}_{n, f, \mathbf{\Gamma}}} \mathbb{P}_{P_S} \left( S > \Gamma + a_n \right) \leq \delta_n^{(3)},
\end{align*}
where $\delta_n^{(3)} \to 0$ as $n \to \infty$. This follows from the uniform upper-tail concentration property in $(\ref{g13})$ and $(\ref{f1})$, which itself follows from Condition \ref{reg_cond2}.

Therefore, we can upper bound 
\begin{align*}
    &\mathbb{E}_{P_S} \left [ \Phi\left( -\frac{\sqrt{2}(N + \Gamma)r}{\sqrt{\Gamma^2 + 2 N S}} + \frac{\sqrt{n}(S - \Gamma)}{\sqrt{2}\sqrt{\Gamma^2 + 2 N S}}  \right)  \right]\\
    &= \int_0^\infty d P_{S}(s) \Phi\left( -\frac{\sqrt{2}(N + \Gamma)r}{\sqrt{\Gamma^2 + 2 N s}} + \frac{\sqrt{n}(s - \Gamma)}{\sqrt{2}\sqrt{\Gamma^2 + 2 N s}}  \right)\\
    &\leq \delta_n^{(1)} + \delta_n^{(2)} +  \delta_n^{(3)} + \mathbb{E}_{P_S} \left [ \Phi\left( -\frac{\sqrt{2}(N + \Gamma)r}{\sqrt{\Gamma^2 + 2 N \Gamma}} + \frac{\sqrt{n}(S - \Gamma)}{\sqrt{2}\sqrt{\Gamma^2 + 2 N \Gamma}}  \right)  \right].
\end{align*}
Hence, 
\begin{align}
    &\min_{ \substack{P_{S} \in \mathcal{S}_{n, f, \mathbf{\Gamma}}\\|\text{supp}(P_S)| \leq k+2} } \mathbb{E}_{P_S} \left [ \Phi\left( \frac{\sqrt{2}(N + \Gamma)r}{\sqrt{\Gamma^2 + 2 N S}} + \frac{\sqrt{n}(\Gamma - S)}{\sqrt{2}\sqrt{\Gamma^2 + 2 N S}}  \right)  \right] \label{bey} \\
    &\geq \min_{ \substack{P_{S} \in \mathcal{S}_{n, f, \mathbf{\Gamma}}\\|\text{supp}(P_S)| \leq k+2} } \mathbb{E}_{P_S} \left [ \Phi\left( \frac{\sqrt{2}(N + \Gamma)r}{\sqrt{\Gamma^2 + 2 N \Gamma}} + \frac{\sqrt{n}( \Gamma - S)}{\sqrt{2}\sqrt{\Gamma^2 + 2 N \Gamma}}  \right)  \right]  - \delta_n^{(1)} - \delta_n^{(2)} -  \delta_n^{(3)}. \notag 
\end{align}
Now with the substitution $U = \sqrt{n}(S - \Gamma)$, we have for every integer $n \geq 1$,
\begin{align*}
    &\min_{ \substack{P_{S} \in \mathcal{S}_{n, f, \mathbf{\Gamma}}\\|\text{supp}(P_S)| \leq k+2} } \mathbb{E}_{P_S} \left [ \Phi\left( \frac{\sqrt{2}(N + \Gamma)r}{\sqrt{\Gamma^2 + 2 N \Gamma}} + \frac{\sqrt{n}( \Gamma - S)}{\sqrt{2}\sqrt{\Gamma^2 + 2 N \Gamma}}  \right)  \right]\\
    &= \inf_{P_{U} \in \mathcal{U}_{f,\mathbf{\Gamma}}^{(k)} } \mathbb{E}_{P_U} \left [  \Phi\left( \frac{\sqrt{2}(N + \Gamma)r}{\sqrt{\Gamma^2 + 2 N \Gamma}} - \frac{U}{\sqrt{2}\sqrt{\Gamma^2 + 2 N \Gamma}}  \right) \right ],
\end{align*}
where $$\mathcal{U}_{f, \mathbf{\Gamma}}^{(k)} \coloneqq \left \{ P \in \mathcal{P}(\mathbb{R}) : |\text{supp}(P_U)| \leq k+2, \mathbb{E}_{P}[U] \leq 0, \mathbb{E}_{P}[f_i(U)] \leq \Gamma_i \text{ for } i = 1, \ldots, k \right \}. $$
Hence, applying the limit as $n \to \infty$ in  $(\ref{bey})$, we obtain 
\begin{align}
    &\liminf_{n \to \infty} \min_{ \substack{P_{S} \in \mathcal{S}_{n, f, \mathbf{\Gamma}}\\|\text{supp}(P_S)| \leq k+2} } \mathbb{E}_{P_S} \left [ \Phi\left( \frac{\sqrt{2}(N + \Gamma)r}{\sqrt{\Gamma^2 + 2 N S}} + \frac{\sqrt{n}(\Gamma - S)}{\sqrt{2}\sqrt{\Gamma^2 + 2 N S}}  \right)  \right] \notag \\
    &\geq \inf_{P_U \in \mathcal{U}_{f, \mathbf{\Gamma}}^{(k)} } \mathbb{E}_{P_U} \left [  \Phi\left( \frac{\sqrt{2}(N + \Gamma)r}{\sqrt{\Gamma^2 + 2 N \Gamma}} - \frac{U}{\sqrt{2}\sqrt{\Gamma^2 + 2 N \Gamma}}  \right) \right ]. \label{n24}
\end{align}
To finish the proof, we take the limit as $n \to \infty$ in $(\ref{b5my})$ and then substitute $(\ref{n24})$ in $(\ref{b5my})$. 

\section{Proof of Lemma \ref{slightlybetterlemma} \label{slightlybetterlemmaproof}}

From Lemma \ref{QccN}, we have 
\begin{align*}
    Q^{cc}(\mathbf{y}) &= \frac{\Gamma\left( \frac{n}{2} \right)}{2 (\pi N)^{n/2}}  \cdot  \exp\left( -\frac{n\Gamma + ||\mathbf{y}||^2}{2N} \right)  \left( \frac{N}{\sqrt{n\Gamma}||\mathbf{y}||}\right)^{\frac{n}{2}-1}  I_{\frac{n}{2}-1}\left(\frac{\sqrt{n\Gamma}||\mathbf{y}||}{N}\right).
\end{align*}
From the standard formula for the multivariate Gaussian, we have 
\begin{align*}
    Q^*(\mathbf{y}) &= \frac{1}{(2\pi(\Gamma' + N))^{n/2}} \exp\left(- \frac{1}{2(\Gamma' + N) } ||\mathbf{y}||^2 \right).
\end{align*}
Then 
\begin{align}
&\log \frac{Q^{cc}(\mathbf{y})}{Q^*(\mathbf{y})} \notag \\
    &= \log \left( \Gamma\left( \frac{n}{2} \right) \right) -  \frac{n\Gamma + ||\mathbf{y}||^2}{2N} + \frac{n}{2} \log\left(  \frac{N}{\sqrt{n\Gamma}||\mathbf{y}||} \right) - \log\left(  \frac{N}{\sqrt{n\Gamma}||\mathbf{y}||} \right) + \log \left( I_{\frac{n}{2}-1}\left(\frac{\sqrt{n\Gamma}||\mathbf{y}||}{N}\right) \right) \notag \\
    & \quad \quad \quad \quad \quad - \log(2) - \frac{n}{2} \log(\pi N) + \frac{n}{2}\log (2\pi(\Gamma' + N)) + \frac{1}{2(\Gamma' + N)} ||\mathbf{y}||^2 \notag \\
    &= \log \left( \Gamma\left( \frac{n}{2} \right) \right) -  \frac{ ||\mathbf{y}||^2 \Gamma' }{2N(\Gamma' + N)} - \frac{n\Gamma}{2N} + \frac{n}{2} \log\left(  \frac{N}{\sqrt{n\Gamma}||\mathbf{y}||} \right) - \log\left(  \frac{N}{\sqrt{n\Gamma}||\mathbf{y}||} \right) + \log \left( I_{\frac{n}{2}-1}\left(\frac{\sqrt{n\Gamma}||\mathbf{y}||}{N}\right) \right) \notag \\
    &  \quad \quad \quad \quad \quad - \log(2)  + \frac{n}{2} \log \left( \frac{2\Gamma' + 2N}{N}\right) \notag  \\
    &\stackrel{(a)}{=} \frac{n}{2} \log\left(\frac{n}{2} \right) - \frac{n}{2} - \frac{1}{2} \log\left(\frac{n}{2} \right) + O(1) -  \frac{ ||\mathbf{y}||^2 \Gamma' }{2N(\Gamma' + N)} - \frac{n\Gamma}{2N} + \frac{n}{2} \log\left(  \frac{N}{\sqrt{n\Gamma}||\mathbf{y}||} \right) - \log\left(  \frac{N}{\sqrt{n\Gamma}||\mathbf{y}||} \right) + \mbox{} \notag \\
	&  \quad \quad \quad \quad \quad   \log \left( I_{\frac{n}{2}-1}\left(\frac{\sqrt{n\Gamma}||\mathbf{y}||}{N}\right) \right)  + \frac{n}{2} \log \left( \frac{2\Gamma' + 2N}{N}\right). \label{b20} 
\end{align}
In equality $(a)$, we used an asymptotic expansion of the log gamma function (see, e.g., \cite[5.11.1]{NISTHandbook}). To approximate the Bessel function, we first rewrite it as  
\begin{align*}
    I_{\frac{n}{2}-1}\left(\frac{\sqrt{n\Gamma}||\mathbf{y}||}{N}\right) &= I_{\nu}\left(\nu z  \right),
\end{align*}
where $\nu = \frac{n}{2} - 1$ and $z = \frac{2  \sqrt{n\Gamma} ||\mathbf{y}||}{N(n-2)}$. Since $\mathbf{y} \in \mathcal{P}_n^*$, we have 
\begin{align}
    z \in \mathcal{Z}_n^* = \left \{z \in \mathbb{R}: \sqrt{\frac{4 \Gamma^2}{N^2}  + \frac{4 \Gamma}{N} + \frac{4\Gamma}{N^2}(\epsilon - \Delta ) + O\left(\frac{1}{n} \right)} \leq z \leq \sqrt{\frac{4 \Gamma^2}{N^2}  + \frac{4 \Gamma}{N} + \frac{4\Gamma}{N^2}(\epsilon  + \Delta)  + O\left(\frac{1}{n} \right)} \right \}.
\end{align}
Since we focus on the $\epsilon \to 0$ asymptotic regime and $\Delta = O(1)$, the variable $z$ can be treated as an $O(1)$ term for the remainder of the proof. In particular, 
$z$ lies in a compact interval $[a, b] \subset (0, \infty)$ for sufficiently large $n$ and small $|\epsilon|$, where $0 < a < b < \infty$. Hence, we can use a uniform asymptotic expansion of the modified Bessel function (see \cite[10.41.3]{NISTHandbook} whose interpretation is given in \cite[2.1(iv)]{NISTHandbook}): as $\nu \to \infty$ and for $0 < z < \infty$, we have 
\begin{align}
    I_{\nu}(\nu z) = \frac{e^{\nu \eta}}{(2 \pi \nu)^{1/2} (1 + z^2)^{\frac{1}{4}}} \left(1+ O \left(\frac{1}{\nu} \right) \right), \label{besselapprox}
\end{align}
where
$$\eta = \sqrt{1 + z^2} + \log \left( \frac{z}{1 + (1 + z^2)^{1/2}}\right)$$
and the $O(1/\nu)$ term in $(\ref{besselapprox})$ can be uniformly bounded over $\mathcal{Z}_n^*$. Using the asymptotic expansion in $(\ref{besselapprox})$ for $\nu = \frac{n}{2} - 1$ and $z = \frac{2\sqrt{n\Gamma}||\mathbf{y}||}{N(n-2)}$, we have 
\begin{align}
    \log I_{\frac{n}{2}-1}\left(\frac{\sqrt{n\Gamma}||\mathbf{y}||}{N}\right) &= \nu \sqrt{1 + z^2} + \nu \log \left( \frac{z}{1 + (1 + z^2)^{1/2}}\right) - \frac{1}{2} \log\left(\nu \right) + O(1)\\
    &= \frac{n}{2} \sqrt{1 + z^2} + \frac{n}{2} \log \left( \frac{z}{1 + (1 + z^2)^{1/2}}\right) - \frac{1}{2} \log\left(n \right) + O(1),  \label{z3}
\end{align}
where it can be verified that the $O(1)$ term can be made to be uniformly bounded over $\mathcal{P}_n^*$ or equivalently $\mathcal{Z}_n^*$. Substituting $(\ref{z3})$ in $(\ref{b20})$ and also substituting $||\mathbf{y}|| = \frac{N(n-2)z}{2 \sqrt{n \Gamma}}$, we obtain  
\begin{align}
&\log \frac{Q^{cc}(\mathbf{y})}{Q^*(\mathbf{y})} \notag \\
    &= \frac{n}{2} \log\left(n \right) - \frac{n}{2} - \frac{1}{2} \log\left(\frac{n}{2} \right)  -  \frac{ N(n-2)^2 z^2 \Gamma' }{8n \Gamma (\Gamma' + N)} - \frac{n\Gamma}{2N} + \frac{n}{2} \log\left(\frac{2}{(n-2)z} \right) - \log\left(\frac{2}{(n-2)z} \right) + \mbox{} \notag \\
    &  \quad \quad \quad \quad \quad   \frac{n}{2} \sqrt{1 + z^2} + \frac{n}{2} \log \left( \frac{z}{1 + (1 + z^2)^{1/2}}\right) - \frac{1}{2} \log\left(n \right)  + \frac{n}{2} \log \left( \frac{\Gamma' + N}{N}\right) + O(1) \notag \\
    &= \frac{n}{2} \log\left(n \right) - \frac{n}{2} - \frac{1}{2} \log\left(\frac{n}{2} \right)  -  \frac{ N n z^2 \Gamma' }{8 \Gamma (\Gamma' + N)} - \frac{n\Gamma}{2N} + \frac{n}{2} \log\left(\frac{2}{(n-2)z} \right) + \log(n) + \mbox{} \notag \\
    &  \quad \quad \quad \quad \quad   \frac{n}{2} \sqrt{1 + z^2} + \frac{n}{2} \log \left( \frac{z}{1 + (1 + z^2)^{1/2}}\right) - \frac{1}{2} \log\left(n \right)  + \frac{n}{2} \log \left( \frac{\Gamma' + N}{N}\right) + O(1) \notag \\
    &=  - \frac{n}{2}   -  \frac{ N n z^2 \Gamma' }{8 \Gamma (\Gamma' + N)} - \frac{n\Gamma}{2N} + \frac{n}{2} \log\left(\frac{2}{z} \right)  +  \frac{n}{2} \sqrt{1 + z^2} + \frac{n}{2} \log \left( \frac{z}{1 + (1 + z^2)^{1/2}}\right)   + \frac{n}{2} \log \left( \frac{\Gamma' + N}{N}\right) + O(1) \notag \\
    &= \frac{n}{2} B + O(1), \label{got7}
\end{align}
where we write 
\begin{align}
    B &=  - 1   -  \frac{ N  z^2 \Gamma' }{4 \Gamma (\Gamma' + N)} - \frac{\Gamma}{N} + \log\left(\frac{2}{z} \right)  +  \sqrt{1 + z^2} +  \log \left( \frac{z}{1 + (1 + z^2)^{1/2}}\right)   + \log \left( \frac{\Gamma' + N}{N}\right). 
\end{align}
We now substitute $\Gamma' = \Gamma + \epsilon$ and do additional simplification to obtain 
\begin{align}
    B &=  - \frac{\Gamma + N}{N}   -  \frac{ N  z^2 (\Gamma + \epsilon) }{4 \Gamma (\Gamma + N + \epsilon)}   +  \sqrt{1 + z^2} +  \log \left( \frac{2}{1 + (1 + z^2)^{1/2}}\right)   + \log \left( \frac{\Gamma + \epsilon + N}{N}\right)\\
    &= - \frac{\Gamma + N}{N}   -  \frac{ N  (\Gamma + \epsilon) }{4 \Gamma (\Gamma + N + \epsilon)}z^2   +  \sqrt{1 + z^2} -  \log \left( \frac{1}{2} + \frac{1}{2} \sqrt{1 + z^2}\right)   + \log \left(1 + \frac{\Gamma}{N} + \frac{\epsilon}{N}\right). 
\end{align}
For convenience, we substitute $s = \sqrt{1 + z^2}$ and write $B = \mu(s, \epsilon)$, where 
\begin{align}
    \mu(s, \epsilon) &= - \frac{\Gamma + N}{N}   -  \frac{ N  (\Gamma + \epsilon) }{4 \Gamma (\Gamma + N + \epsilon)}(s^2 - 1)   +  s -  \log \left( \frac{1}{2} + \frac{1}{2} s\right)   + \log \left(1 + \frac{\Gamma}{N} + \frac{\epsilon}{N}\right). \label{musi}
\end{align}
Note that 
\begin{align}
    s \in \mathcal{S}_n^* = \left \{s \in \mathbb{R}: \sqrt{\frac{4\Gamma^2}{N^2} + \frac{4\Gamma}{N} + \frac{4\Gamma}{N^2}(\epsilon - \Delta) + O\left(\frac{1}{n} \right) + 1} \leq s \leq \sqrt{\frac{4\Gamma^2}{N^2} + \frac{4\Gamma}{N} + \frac{4\Gamma}{N^2}(\epsilon + \Delta) +O\left(\frac{1}{n} \right)+ 1} \right \}. \label{Sdomain}
\end{align}
Define 
\begin{align}
    F(\epsilon) = \sup_{s \in \mathcal{S}_n^*} \mu(s, \epsilon).  \label{maxovers}
\end{align}
We have 
\begin{align}
    \frac{\partial \mu (s, \epsilon) }{\partial s} = \frac{s ( \Gamma N + 2 \Gamma^2 + 2 \Gamma \epsilon  - (\Gamma N + N \epsilon)s  - N \epsilon )}{2 \Gamma  (s+1) (\Gamma +N+\epsilon )},
\end{align}
which is nonnegative for $s < s^*(\epsilon)$ and nonpositive for
$s > s^*(\epsilon)$, where
\begin{align}
    s^*(\epsilon) &=  \frac{\Gamma N + 2 \Gamma^2 + 2 \Gamma \epsilon - N \epsilon}{N(\Gamma + \epsilon)} . 
\end{align}
Thus $s^*(\epsilon)$ is a global maximizer.
We also have that $s^*(\epsilon) \to 1 + \frac{2 \Gamma}{N}$ as $\epsilon \to 0$, and 
it can be checked that for any $\Delta > 0$ and sufficiently small $\epsilon$, $s^*(\epsilon) \in \mathcal{S}_n^*$. 
Hence, 
\begin{align*}
    F(\epsilon) &= \mu(s^*(\epsilon), \epsilon)\\
    &= - \frac{\epsilon}{\Gamma + \epsilon} + \log\left(1 + \frac{\epsilon}{\Gamma} \right)\\
    &= \frac{\epsilon ^2}{2 \Gamma ^2}-\frac{2 \epsilon ^3}{3 \Gamma ^3} + O(\epsilon^4). 
\end{align*}
It follows that
\begin{align}
    B \leq \frac{\epsilon ^2}{2 \Gamma ^2}-\frac{2 \epsilon ^3}{3 \Gamma ^3} + O(\epsilon^4). \label{bupp}
\end{align} 
Substituting $(\ref{bupp})$ in $(\ref{got7})$, we obtain 
\begin{align}
    \log \frac{Q^{cc}(\mathbf{y})}{Q^*(\mathbf{y})} &\leq \frac{n\epsilon ^2}{4 \Gamma ^2}-\frac{n \epsilon ^3}{3 \Gamma ^3} + O(n\epsilon^4) + O\left(1 \right). 
\end{align}
The big $O$ terms are independent of $\mathbf{y}$ since the dependence on $\mathbf{y}$ was only through $s$, which we maximized over in $(\ref{maxovers})$.

\section*{Acknowledgment}

This research was supported by the US National Science
Foundation under grant CCF-1956192.

\bibliographystyle{IEEEtran}
%



\bibliography{citations}

@ARTICLE{dytso2019,
  author={Dytso, Alex and Al, Mert and Poor, H. Vincent and Shamai Shitz, Shlomo},
  journal={IEEE Transactions on Information Theory}, 
  title={On the Capacity of the Peak Power Constrained Vector {G}aussian Channel: An Estimation Theoretic Perspective}, 
  year={2019},
  volume={65},
  number={6},
  pages={3907-3921},
  doi={10.1109/TIT.2018.2890208}}

@ARTICLE{dytso2020,
  author={Dytso, Alex and Yagli, Semih and Poor, H. Vincent and Shamai Shitz, Shlomo},
  journal={IEEE Transactions on Information Theory}, 
  title={The Capacity Achieving Distribution for the Amplitude Constrained Additive {G}aussian Channel: An Upper Bound on the Number of Mass Points}, 
  year={2020},
  volume={66},
  number={4},
  pages={2006-2022},
  doi={10.1109/TIT.2019.2948636}}

@ARTICLE{8012458,
  author={Fong, Silas L. and Tan, Vincent Y. F.},
  journal={IEEE Trans.\ Inf.\ Theory}, 
  title={A Tight Upper Bound on the Second-Order Coding Rate of the Parallel {G}aussian Channel With Feedback}, 
  year={2017},
  volume={63},
  number={10},
  pages={6474-6486},
  doi={10.1109/TIT.2017.2740956}}

@book{Cover2006,
  author    = {T. M. Cover and J. A. Thomas},
  title     = {Elements of Information Theory},
  edition   = {2nd},
  publisher = {Wiley-Interscience},
  address   = {Hoboken, N.J.},
  year      = {2006}
}

@ARTICLE{mahmood2025channelcodinggaussianchannels,
  author={Mahmood, Adeel and Wagner, Aaron B.},
  journal={IEEE Transactions on Information Theory}, 
  title={Channel Coding for {G}aussian Channels With Mean and Variance Constraints}, 
  year={2025},
  volume={71},
  number={12},
  pages={9285-9301},
  doi={10.1109/TIT.2025.3614670}}

@InProceedings{Altug:ISIT10,
  author =       "Y{\"{u}}cel~Altu\u{g} and Aaron~B.~Wagner",
  title =        "Moderate Deviation Analysis of Channel Coding:
                              Discrete Memoryless Case",
  booktitle =    "Proc. IEEE Intl.\ Symp.\ Inf.\ Theory",
  month =        jun,
  year =         "2010",
  pages =        "265--269",
}

@ARTICLE{mahmood2024improvedchannelcodingperformance,
  author={Mahmood, Adeel and Wagner, Aaron B.},
  journal={IEEE Transactions on Communications}, 
  title={Improved Channel Coding Performance Through Cost Variability}, 
  year={2025},
  volume={73},
  number={11},
  pages={10145-10155},
  doi={10.1109/TCOMM.2025.3588600}}

@ARTICLE{mahmood2024channelcodingmeanvariance,
  author={Mahmood, Adeel and Wagner, Aaron B.},
  journal={IEEE Trans.\ Inf.\ Theory}, 
  title={Channel Coding With Mean and Variance Cost Constraints}, 
  year={2025},
  volume={71},
  number={3},
  pages={1504-1532},
  doi={10.1109/TIT.2025.3533947}}

@article{winkler1988extreme,
  author  = {Gerhard Winkler},
  title   = {Extreme Points of Moment Sets},
  journal = {Mathematics of Operations Research},
  volume  = {13},
  number  = {4},
  pages   = {581--587},
  year    = {1988},
  doi     = {10.1287/moor.13.4.581},
  url     = {http://www.jstor.org/stable/3689944}
}

@ARTICLE{kostina_JSCC,
  author={Kostina, Victoria and Verdú, Sergio},
  journal={IEEE Trans.\ Inf.\ Theory}, 
  title={Lossy Joint Source-Channel Coding in the Finite Blocklength Regime}, 
  year={2013},
  volume={59},
  number={5},
  pages={2545-2575},
  doi={10.1109/TIT.2013.2238657}}

@ARTICLE{Marton:Exponent,
  author={Marton, K.},
  journal={IEEE Trans.\ Inf.\ Theory}, 
  title={Error exponent for source coding with a fidelity criterion}, 
  year={1974},
  volume={20},
  number={2},
  pages={197-199},
  doi={10.1109/TIT.1974.1055204}}

@ARTICLE{Shi:Semantic,
  author={Shi, Yuxuan and Shao, Shuo and Wu, Yongpeng and Zhang, Wenjun and Xia, Xiang-Gen and Xiao, Chengshan},
  journal={IEEE Transactions on Wireless Communications}, 
  title={Excess Distortion Exponent Analysis for Semantic-Aware {MIMO} Communication Systems}, 
  year={2023},
  volume={22},
  number={9},
  pages={5927-5940},
  doi={10.1109/TWC.2023.3238463}}

@ARTICLE{Zhou:Two,
  author={Zhou, Lin and Motani, Mehul},
  journal={IEEE Trans.\ Inf.\ Theory}, 
  title={Non-Asymptotic Converse Bounds and Refined Asymptotics for Two Source Coding Problems}, 
  year={2019},
  volume={65},
  number={10},
  pages={6414-6440},
  doi={10.1109/TIT.2019.2920893}}

@ARTICLE{Truong:Joint:Feedback,
  author={Truong, Lan V. and Tan, Vincent Y. F.},
  journal={IEEE Trans.\ Inf.\ Theory}, 
  title={The Reliability Function of Variable-Length Lossy Joint Source-Channel Coding With Feedback}, 
  year={2019},
  volume={65},
  number={8},
  pages={5028-5042},
  doi={10.1109/TIT.2019.2911527}}

@ARTICLE{Venkataramanan:SPARC:Excess,
  author={Venkataramanan, Ramji and Tatikonda, Sekhar},
  journal={IEEE Trans.\ Inf.\ Theory}, 
  title={The Rate-Distortion Function and Excess-Distortion Exponent of Sparse Regression Codes With Optimal Encoding}, 
  year={2017},
  volume={63},
  number={8},
  pages={5228-5243},
  doi={10.1109/TIT.2017.2716360}}

@ARTICLE{Venkataramanan:Feedforward,
  author={Venkataramanan, Ramji and Pradhan, S. Sandeep},
  journal={IEEE Trans.\ Inf.\ Theory}, 
  title={Source Coding With Feed-Forward: Rate-Distortion Theorems and Error Exponents for a General Source}, 
  year={2007},
  volume={53},
  number={6},
  pages={2154-2179},
  doi={10.1109/TIT.2007.896887}}

@ARTICLE{Zhou:SR,
  author={Zhou, Lin and Tan, Vincent Y. F. and Motani, Mehul},
  journal={IEEE Trans.\ Inf.\ Theory}, 
  title={Second-Order and Moderate Deviations Asymptotics for Successive Refinement}, 
  year={2017},
  volume={63},
  number={5},
  pages={2896-2921},
  doi={10.1109/TIT.2017.2674675}}

@ARTICLE{Zhou:GrayWyner,
  author={Zhou, Lin and Tan, Vincent Y. F. and Motani, Mehul},
  journal={IEEE Trans.\ Inf.\ Theory}, 
  title={Discrete Lossy {G}ray–{W}yner Revisited: Second-Order Asymptotics, Large and Moderate Deviations}, 
  year={2017},
  volume={63},
  number={3},
  pages={1766-1791},
  doi={10.1109/TIT.2016.2644670}}

@ARTICLE{Kelly:WZ,
  author={Kelly, Benjamin G. and Wagner, Aaron B.},
  journal={IEEE Trans.\ Inf.\ Theory}, 
  title={Reliability in Source Coding With Side Information}, 
  year={2012},
  volume={58},
  number={8},
  pages={5086-5111},
  doi={10.1109/TIT.2012.2201346}}

@ARTICLE{Zhong:Continuous,
  author={Zhong, Yangfan and Alajaji, Fady and Campbell, L. Lorne},
  journal={IEEE Trans.\ Inf.\ Theory}, 
  title={Joint Source–Channel Coding Excess Distortion Exponent for Some Memoryless Continuous-Alphabet Systems}, 
  year={2009},
  volume={55},
  number={3},
  pages={1296-1319},
  doi={10.1109/TIT.2008.2011436}}

@ARTICLE{Zhong:Joint,
  author={Yangfan Zhong and Alajaji, F. and Campbell, L.L.},
  journal={IEEE Trans.\ Inf.\ Theory}, 
  title={On the joint source-channel coding error exponent for discrete memoryless systems}, 
  year={2006},
  volume={52},
  number={4},
  pages={1450-1468},
  doi={10.1109/TIT.2006.871608}}

@ARTICLE{Weissman:Exponent:Universal,
  author={Weissman, T.},
  journal={IEEE Trans.\ Inf.\ Theory}, 
  title={Universally attainable error exponents for rate-distortion coding of noisy sources}, 
  year={2004},
  volume={50},
  number={6},
  pages={1229-1246},
  doi={10.1109/TIT.2004.828061}}

@ARTICLE{Iriyama:General,
  author={Iriyama, K.},
  journal={IEEE Trans.\ Inf.\ Theory}, 
  title={Probability of error for the fixed-length lossy coding of general sources}, 
  year={2005},
  volume={51},
  number={4},
  pages={1498-1507},
  doi={10.1109/TIT.2004.842777}}

@ARTICLE{Hen:Trellis,
  author={Hen, I. and Merhav, N.},
  journal={IEEE Trans.\ Inf.\ Theory}, 
  title={On the error exponent of trellis source coding}, 
  year={2005},
  volume={51},
  number={11},
  pages={3734-3741},
  doi={10.1109/TIT.2005.856953}}

@ARTICLE{9099482,  author={Wagner, Aaron B. and Shende, Nirmal V. and Altuğ, Yücel},  journal={IEEE Trans.\ Inf.\ Theory},   title={A New Method for Employing Feedback to Improve Coding Performance},   year={2020},  volume={66},  number={11},  pages={6660-6681},  doi={10.1109/TIT.2020.2997385}}

@ARTICLE{Altug:MDP,
  author={Altuğ, Yücel and Wagner, Aaron B.},
  journal={IEEE Trans.\ Inf.\ Theory}, 
  title={Moderate Deviations in Channel Coding}, 
  year={2014},
  volume={60},
  number={8},
  pages={4417-4426},
  doi={10.1109/TIT.2014.2323418}}

@ARTICLE{6145679,
  author={Kostina, Victoria and Verd\'{u}, Sergio},
  journal={IEEE Trans.\ Inf.\ Theory}, 
  title={Fixed-Length Lossy Compression in the Finite Blocklength Regime}, 
  year={2012},
  volume={58},
  number={6},
  pages={3309-3338},
  doi={10.1109/TIT.2012.2186786}}

@book{AliprantisBorder2006,
  author       = {Aliprantis, Charalambos D. and Border, Kim C.},
  title        = {Infinite Dimensional Analysis: A Hitchhiker's Guide},
  edition      = {3},
  year         = {2006},
  publisher    = {Springer},
  address      = {Berlin Heidelberg},
  pages        = {298--299},
  note         = {See Theorem 7.69 (Bauer Maximum Principle)},
  isbn         = {978-3-540-26316-9}
}

@article{esseen11,
  author = {van Beek, P.},
  title = {An application of {F}ourier methods to the problem of sharpening the {B}erry-{E}sseen inequality},
  journal = {Zeitschrift für Wahrscheinlichkeitstheorie und Verwandte Gebiete},
  volume = {23},
  number = {3},
  pages = {187-196},
  year = {1972},
  doi = {10.1007/BF00536558}
}

@ARTICLE{7055296,
  author={Kostina, Victoria and Verdú, Sergio},
  journal={IEEE Trans.\ Inf.\ Theory}, 
  title={Channels With Cost Constraints: Strong Converse and Dispersion}, 
  year={2015},
  volume={61},
  number={5},
  pages={2415-2429},
  doi={10.1109/TIT.2015.2409261}}

@ARTICLE{5290292,
  author={Hayashi, Masahito},
  journal={IEEE Trans.\ Inf.\ Theory}, 
  title={Information Spectrum Approach to Second-Order Coding Rate in Channel Coding}, 
  year={2009},
  volume={55},
  number={11},
  pages={4947-4966},
  doi={10.1109/TIT.2009.2030478}}

@INPROCEEDINGS{strassen,
author = {V. Strassen},
title = {Asymptotische Abschätzungen in {S}hannon's Informationstheorie},
booktitle = {Proc. Trans. 3rd Prague Conf Inf. Theory},
year = {1962},
address = {Prague, Czech},
pages = {689--723}
}

@phdthesis{Polyanskiy2010,
  author    = {Y. Polyanskiy},
  title     = {Channel coding: Non-asymptotic fundamental limits},
  school    = {Dept. Elect. Eng., Princeton Univ., Princeton, NJ, USA},
  year      = {2010},
}

@ARTICLE{7156144,
  author={Yang, Wei and Caire, Giuseppe and Durisi, Giuseppe and Polyanskiy, Yury},
  journal={IEEE Trans.\ Inf.\ Theory}, 
  title={Optimum Power Control at Finite Blocklength}, 
  year={2015},
  volume={61},
  number={9},
  pages={4598-4615},
  doi={10.1109/TIT.2015.2456175}}

@ARTICLE{7300429,
  author={MolavianJazi, Ebrahim and Laneman, J. Nicholas},
  journal={IEEE Trans.\ Inf.\ Theory}, 
  title={A Second-Order Achievable Rate Region for {G}aussian Multi-Access Channels via a Central Limit Theorem for Functions}, 
  year={2015},
  volume={61},
  number={12},
  pages={6719-6733},
  doi={10.1109/TIT.2015.2492547}}

@ARTICLE{5452208,
  author={Polyanskiy, Yury and Poor, H. Vincent and Verd\'{u}, Sergio},
  journal={IEEE Trans.\ Inf.\ Theory}, 
  title={Channel Coding Rate in the Finite Blocklength Regime}, 
  year={2010},
  volume={56},
  number={5},
  pages={2307-2359},
  doi={10.1109/TIT.2010.2043769}}

@INPROCEEDINGS{review12,
  author={Kim, Young-Han and Lapidoth, Amos and Weissman, Tsachy},
  booktitle={2010 IEEE Information Theory Workshop on Information Theory (ITW 2010, Cairo)}, 
  title={Error exponents for the {G}aussian channel with noisy active feedback}, 
  year={2010},
  volume={},
  number={},
  pages={1-3},
  doi={10.1109/ITWKSPS.2010.5503226}}

@ARTICLE{4thorderkostina,
  author={Yavas, Recep Can and Kostina, Victoria and Effros, Michelle},
  journal={IEEE Transactions on Information Theory}, 
  title={Third-Order Analysis of Channel Coding in the Small-to-Moderate Deviations Regime}, 
  year={2024},
  volume={70},
  number={9},
  pages={6139-6170},
  doi={10.1109/TIT.2024.3426509}}

@ARTICLE{review13,
  author={Vazquez-Vilar, Gonzalo},
  journal={IEEE Transactions on Information Theory}, 
  title={Error Probability Bounds for {G}aussian Channels Under Maximal and Average Power Constraints}, 
  year={2021},
  volume={67},
  number={6},
  pages={3965-3985},
  doi={10.1109/TIT.2021.3063311}}

@ARTICLE{review11,
  author={Shannon, Claude E.},
  journal={The Bell System Technical Journal}, 
  title={Probability of error for optimal codes in a {G}aussian channel}, 
  year={1959},
  volume={38},
  number={3},
  pages={611-656},
  doi={10.1002/j.1538-7305.1959.tb03905.x}}

@ARTICLE{shaped_QAM,
  author={Fehenberger, Tobias and Alvarado, Alex and Böcherer, Georg and Hanik, Norbert},
  journal={Journal of Lightwave Technology}, 
  title={On Probabilistic Shaping of Quadrature Amplitude Modulation for the Nonlinear Fiber Channel}, 
  year={2016},
  volume={34},
  number={21},
  pages={5063-5073},
  doi={10.1109/JLT.2016.2594271}}

@ARTICLE{7056434,
  author={Tan, Vincent Yan Fu and Tomamichel, Marco},
  journal={IEEE Trans.\ Inf.\ Theory}, 
  title={The Third-Order Term in the Normal Approximation for the {AWGN} Channel}, 
  year={2015},
  volume={61},
  number={5},
  pages={2430-2438},
  doi={10.1109/TIT.2015.2411256}}

@book{NISTHandbook,
  title={NIST Handbook of Mathematical Functions},
  editor={Olver, Frank W.~J. and Lozier, Daniel W. and Boisvert, Ronald F. and Clark, Charles W.},
  year={2010},
  publisher={Cambridge University Press},
  address={New York},
  isbn={9780521140638}
}

@Book{Billingsley:Convergence2,
  author =       "Patrick~Billingsley",
  title =        "Convergence of Probability Measures",
  publisher =    "John Wiley \& Sons",
  address =      "New York",
  year =         "1999",
  edition =      "2nd",
}

@Book{Chung:Probability3,
  author =       "Kai Lai Chung",
  title =        "A Course in Probability",
  publisher =    "Academic Press",
  address =      "San Diego",
  year =         "2001",
  edition =      "3rd",
}

@ARTICLE{Hori:Overrun,
  author={Masaki Hori and  Mikihiko Nishiara},
  journal={IEICE Trans.\ on Fund.\ Electr., Comm., and Comp.\ Sci.},
  title={Channel Capacity with Cost Constraint Allowing Cost Overrun},
  year={2024},
  volume={E107-A},
  number={3},
  pages={458-463},
  doi={10.1587/transfun.2023TAP0010},
}

@ARTICLE{Nishiara:Delivery,
  author={Mikihiko Nishiara},
  journal={IEICE Trans.\ on Fund.\ Electr., Comm., and Comp.\ Sci.},
  title={Channel Coding with Cost Paid on Delivery},
  year={2022},
  volume={E105-A},
  number={3},
  pages={345-352},
  doi={10.1587/transfun.2021TAP0002},
}

@INPROCEEDINGS{Hori:Overrun:ISITA,
  author={Hori, Masaki and Nishiara, Mikihiko},
  booktitle={Intl.\ Symp.\ on Inf.\ Theory and Its Appl.\ (ISITA)}, 
  title={Channel Capacity with Cost Constraint Allowing Some Cost Overrun}, 
  year={2022},
  pages={34-38},
}

\end{document}